# Geological Mapping and Chronology of Lunar Landing Sites: Apollo 11

W. Iqbal[1], H. Hiesinger[1], and C. H. van der Bogert[1]

[1]Institut für Planetologie, Westfälische Wilhelms-Universität Münster

Wilhelm-Klemm-Str. 10, 48149 Münster, Germany

Tel.: +49 251 8339075

Fax: +49 251 8336301

iqbalw@uni-muenster.de

hiesinger@uni-muenster.de

vanderbogert@uni-muenster.de

# Abstract

Crater size-frequency distribution (CSFDs) measurements allow the derivation of absolute model ages (AMAs) for geological units across various terrestrial bodies in the Solar System based on body-specific adjustments to the lunar chronology (e.g., Hartmann, 1970; Neukum et al., 1975, 1983, 2001; Stöffler et al., 2001, 2006; Hiesinger et al., 2012; Robbins, 2014). Thus, it is important to revisit and test the accuracy of the lunar chronology using data from recent lunar missions (e.g., Hiesinger et al., 2000, 2012, 2015; Rajmon and Spudis, 2004; Stöffler et al., 2006), as well as newer analyses of lunar samples (e.g., Gaffney et al. 2011, Meyer, 2012;

Snape et al., 2016; Welsh et al., 2018). We generated a new detailed geological map of the Apollo 11 landing region based on spectral characteristics, topography, and albedo maps, which shows several mare units adjacent to the lunar module. Lunar Reconnaissance Orbiter Camera (LROC) images were used to measure new CSFDs and derive the cumulative number of craters with diameters ≥1 km or $N(1)$ for the Apollo 11 landing site. The newly derived $N(1)$ values are consistent with the presence of only one surficial unit at the landing site: the Group A, High-K (high potassium) "young" mare basalt (Meyer, 2012). We reviewed the radiometric ages for Apollo 11 samples that have been determined since the calibration of the lunar cratering chronology, used our new geological map to reinterpret their provenance, and correlated them with the new $N(1)$ values. These are plotted and compared with the lunar chronology of Neukum et al. (1983). Our calibration point for the Apollo 11 landing site is consistent with the earlier values, thus, confirming Neukum's (1983) lunar chronology curve.

# 1. Introduction

The relative ages of geological features on planetary surfaces are often determined via the comparison of stratigraphic relationships. In addition, crater spatial densities of particular areas can also provide relative age information, because regions with more craters have an older age than regions with fewer craters. Crater size-frequency distribution (CSFD) measurements of relative crater spatial densities of geological units at the Apollo and Luna landing sites were calibrated to the radiometric and exposure ages of samples collected at the sites, thus, allowing the determination of absolute model ages (AMAs) for geological units across the Moon and on planetary bodies throughout the Solar System (e.g., Hartmann, 1970; Neukum et al., 1975, 1983, 2001; Ivanov et al., 2001, 2002; Hiesinger et al., 2000, 2002, 2012; Stöffler et al., 2001, 2006; Robbins, 2014).

The Apollo 11 landing site is an important calibration point for the lunar chronology because of the well-studied radiometric and exposure ages of the returned samples (Kramer et al., 1977; Beaty and Albee, 1978; Guggisberg et al., 1979; Snyder et al., 1996; Stöffler et al., 2006; Gaffney et al. 2011; Meyer, 2012; Snape et al., 2016). The landing site is located in Mare Tranquillitatis, between rays from the relatively young craters, Theophilus, Alfraganus, and Tycho (Aldrin et al., 1969; Grolier, 1970; Stöffler et al. 2006) (Fig. 1). Detailed studies of the samples showed five different chemical classes of mare basalts, representing four groups of radiometric ages: Group A (3.58 Ga), a high potassium basalt; Group B1-B3 (3.70 Ga), a complex group; and the two oldest groups Group B2 (3.80 Ga), and D (3.85 Ga) (Stöffler et al., 2006; Meyer, 2012; Welsh et al., 2018). Despite the chemical differences between these basalts, they are all rich in titanium (Papanastassiou et al., 1970; Beaty and Albee, 1980; Snyder et al., 1995; Stöffler et al., 2006). Clementine data revealed various mare units on the basis of spectral differences across Mare Tranquillitatis, which supports the identification of different basalt groups in the returned samples (Pieters et al., 1994; Rajmon and Spudis, 2004).

Neukum (1983) derived two calibration points for the lunar chronology from the Apollo 11 landing site. Neukum and coworkers (e.g., Neukum and Ivanov, 1994) observed different crater spatial density units representing the presence of different lava flows and inferred that the young one is the 3.58 Ga old group A basalts and the older one is the 3.80 Ga old group B2 or D basalts. Robbins (2014) measured the CSFDs in a 10x larger region of Mare Tranquillitatis around the Apollo 11 landing site, and suggests that slightly larger $N(1)$ value represents the "old" basalt unit. Given that older basalts are present in the sample collection, it is likely that the larger Robbins (2014) area includes these older basalts. Neukum (1983) selected smaller, local measurement areas, because he argued that the count areas should be restricted to locations at the landing sites that clearly represent individual sampled geological units.

We used recent data, including multispectral (Pieters et al., 1994), topographic (Scholten et al., 2012; Barker et al., 2016), and high-resolution image data with various illumination geometries (Robinson et al., 2010) for a new detailed geological investigation of the Apollo 11 landing site, in order to check and improve the derivation of AMAs via CSFD measurements, and compare the relative crater frequencies with recently updated and measured lunar sample ages (e.g., Stöffler et al., 2006; Meyer, 2012; Snape et al., 2016). We include an appendix with reference images and individual CSFD measurement plots.

# 2. Methodology

## 2.1 Data

For the geological mapping of the Apollo 11 landing site and the surrounding region, we used the LRO WAC (Fig. 1a) and NAC albedo images (Fig. 2) (Robinson et al., 2010), the LRO WAC digital terrain model (DTM) (Scholten et al., 2012), the Lunar Orbiter Laser Altimeter (LOLA) and SELENE Terrain camera merged digital elevation model (DEM) (Barker et al., 2016), and the Clementine UV-VIS color ratio mosaic (Pieters et al., 1994). Major geological structures were identified with the LOLA-SELENE Terrain Camera DEM (Barker et al., 2016), which has a pixel scale of 59 m. We also used this data as a base map. The LRO WAC DTM has a pixel scale of 236 m and shows that the lowest elevation is -4.2 km and the highest elevation is 1.7 km in the study area. The raw LRO data were calibrated and map-projected with ISIS3 (Anderson et al., 2004), and the processed data was obtained from the team products of the LRO Science Operation Center (Henriksen et al., 2016). Data currently unavailable publicly at the time of publication will become available via regular Planetary Data System releases.

The Clementine UVVIS camera (Fig. 1b) data have a pixel scale of 100 m (Pieters et al., 1994; Rajmond and Spudis, 2004). Ratios of monochromatic mosaics: R = 750/415, G = 750/950, B = 415/750 were used to create a false-color image from the Clementine UVVIS data (Rajmond and Spudis, 2004). The color ratios exaggerate differences in the compositional and maturity levels of the geological units. Generally, in Clementine false-color mosaic, red represents mature highland material, light to medium blue excavated highland, orange low-Ti basalts, dark blue high-Ti basalts, and yellow to green excavated basalts (Fig. 1b and also see Appendix Fig. A1) (Rajmond and Spudis, 2004). Different mare units were identified in Clementine data primarily due to their spectral differences caused by the varying concentrations of iron and titanium in the overlying regolith (Fig. 4).

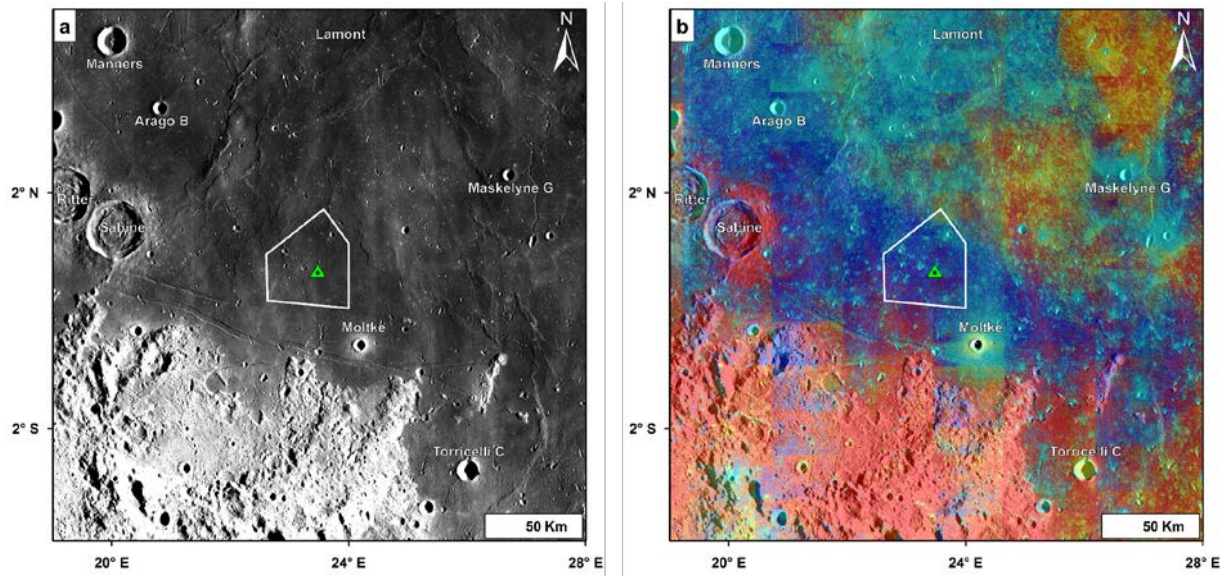

*Fig. 1: The Apollo 11 landing site (green triangle), located in Mare Tranquillitatis, was studied using (a) the LRO WAC mosaic and (b) the Clementine UV-VIS false color ratio mosaic, to map different mare units on the basis of spectral differences. The white polygon around the landing site shows the extent of the original count area of Neukum (1983), which was used to measure CSFDs for the Apollo 11 landing site.*

For CSFD measurements on LRO NAC data, we used image pairs M102000149 (pixel scale 1.19 m; incidence angle 79.5°), M150361817 (pixel scale 0.5 m; incidence angle 62.5°), and M162154734 (pixel scale 0.5 m; incidence angle 79.5°). Along with these image pairs, we used LRO NAC-derived controlled mosaics from the image pair M16161085 (pixel scale 0.5 m; incidence angle 81.7°), which has an overall pixel scale of 1.81 m, (Fig. 2a) and a DTM mosaic of image pairs M150361817 and M150368601 (Fig. 2b) (Henriksen et al., 2016). The NAC DTM shows a lowest elevation of -2017 m and a highest elevation of -1808 m. An orthographic projection (Longitude of the center: 23.6 and Latitude of the Center 0.69) was used for our study to minimize distortions (Anderson et al., 2004; Robinson et al., 2010).

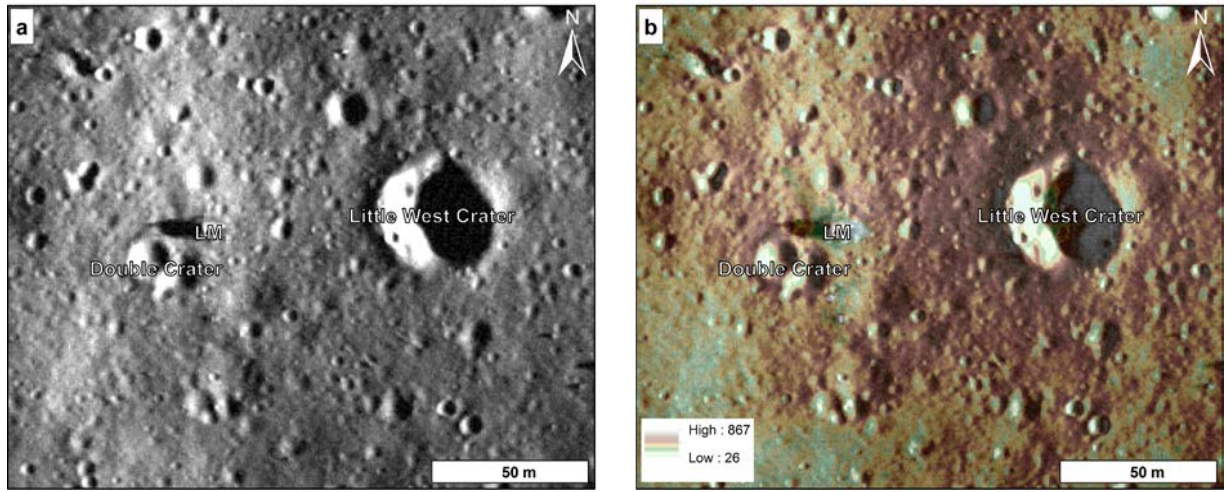

*Fig. 2: LRO NAC data was used for the detailed study of the landing site, as well as CSFD measurements. These data include: (a) the controlled mosaic of image pair M16161085 and (b) the NAC DTM mosaic of image pair M150361817.*

## 2.2 Mapping Technique

Our detailed geological map is based on a detailed inspection of albedo, spectral, and topographic information. For the mapping of the mare units in southwestern Mare Tranquillitatis, we applied the same approach of hybrid spectral mapping on the Clementine data as Hiesinger et al. (2000). The overall stratigraphic scheme for the mapping was adapted from Fortezzo and Hare (2013), who produced a digital version of various geological maps at 1:5M scale. Fortezzo and Hare (2013) followed the stratigraphic scheme proposed by Wilhelms (1987). The symbology used for mapping the geological features follows the standards of the Federal Geographic Data Committee (2006). The nomenclature for the mapped craters and regions were taken from the Gazetteer of Planetary Nomenclature (Blue, 1999)

Major differences in the geological units were observed through albedo contrast (here used as a qualitative characteristic) on LRO WAC data (Robinson et al., 2010). For example, the high albedo highland material shows a clear contrast from the low albedo mare units. The extent of the ray material from the young Copernican craters was also noted as an albedo contrast. However, different mare unit boundaries were observed and defined using the variation in FeO and $TiO_2$ content as seen in the Clementine data (Rajmond and Spudis, 2004). The extents of the geological structures, including normal and reverse faults, crater rims, kipukas, and other elevated structures were marked using the LOLA-SELENE DEM (in meters) as a reference (Barker et al., 2016).

## 2.3 CSFDs Measurement Technique and Lunar Cratering Chronology

The technique to determine relative and absolute model ages from CSFD measurements involves two basic steps: (1) identification and measurement of a homogeneous geologic area and (2) accurate measurement of the numbers and diameters of primary craters within this area (e.g., Hartmann, 1970; Neukum et al., 1975, 1983, 2001; Hiesinger et al., 2012). Traditionally, the obtained crater diameters are distributed into diameter bins, specified, for example in Crater Analysis Techniques Working Group (1979), Neukum et al. (2001), Stöffler et al. (2006), Kneissl et al. (2011), and Michael et al. (2010, 2013, 2016). For our study, we used pseudolog binning and cumulative plotting and fitting of our data (Michael et al., 2010, 2013, and 2016) to allow it to be directly compared with previous work by Neukum et al. (2001) and Hiesinger et al. (2011, 2012). The geological units of different ages contain crater distributions that have the same shape in overlapping diameter ranges (e.g., Hartmann, 1970; Neukum et al., 1975, 1983, 2001; Marchi, 2009). These distributions can be aligned along a complex continuous curve, called the lunar production function (Neukum et al. 1975, 1983, 1994 and 2001), by shifting them in log $N_{cum}$ i.e. vertical direction (Neukum et al. 1975, 1983, 1994 and 2001; Hiesinger et al. 2000). Hence, the production function (Neukum et al. 1983, 2001) is a polynomial function, defining the crater size-frequency in diameter range for a certain time of exposure. Neukum et al (1983. 2001) derived the lunar production function as an eleventh-degree polynomial function:

$$Log(N_{cum}) = a_0 + \sum_{k=1}^{11} a_k [log(D)]^k \qquad (1)$$

Here, $a_0$ represents the distribution density (Michael et al. 2016) and D represents the diameter range of certain size. The production function of Neukum et al. (2001) has been independently tested (e.g., Hiesinger et al., 2012) and shown to be a good fit for the craters in the diameter range of 10 m to 100 km.

The lunar cratering chronology was established (e.g., Neukum et al., 1983, 2001) by correlating the radiometric/exposure ages of Apollo and Luna samples to the cumulative number of craters greater than or equal to a reference diameter within the geological unit. This correlation defines the impact rate on the moon as a function of time represented as lunar cratering chronology curve (e.g., Neukum et al., 1983, 2001). The calculation of the chronology is well described in

e,g., Hartmann. (1970), Neukum et al. (1975, 1983, 2001), Stöffler et al. (2001, 2006) and Hiesinger et al., (2000). The chronology function relates the accumulated distribution density to the surface exposure age. The arbitrary reference diameter for which cumulative density can be given is 1 km. The chronology function is expressed analytically as:

$$N(1) = 5.44 \times 10^{-14}[\exp(6.93T)-1] + 8.38 \cdot 10^{-4}T \qquad (2)$$

Equation 2 correlates craters of a certain diameter range in $km^2$ area crater accumulation (retention) age **T** in Ga (billion years).

To evaluate potential contamination from secondary craters during the measurement of the crater diameters, Michael et al. (2012) developed a randomness analysis to aid in identifying crater clustering on different scales through the mean closest neighbor (crater) distance in the area. This analysis identifies diameter ranges of craters that show clusters and chains. Thus, the error related to measured *N*(1) values can be reduced by avoiding contamination from unnoticeable secondary craters.

We measured CSFDs for the Apollo 11 landing site and its vicinity on Lunar Reconnaissance Orbiter Wide Angle (WAC) mosaic and Narrow Angle Camera (NAC) images. After processing, the images were imported into ArcGIS, and the extension CraterTools (Kneissl et al., 2011) was used to perform CSFD measurements of homogeneous geologic areas. The areas used for measuring the *N*(1) values were carefully selected to avoid obvious secondary craters chains and clusters (McEwen and Bierhaus, 2006), on the basis of their morphology. Few clusters and chains were also avoided by excluding the marked diameters of these clusters during the CSFD measurements. Afterward, the randomness analysis (Michael et al., 2012) was applied to identify clusters or chains of the secondary craters to minimize their effects on our derived AMAs. Crater diameters smaller than 50 m show some degree of clustering in the randomness analysis (Michael et al., 2012). However, these craters were not used to derive *N*(1) values and, thus, do not affect the accuracy of our results.

On WAC images, we measured the same area as Neukum (1983), which was used for the derivation of the original lunar chronology of Neukum (1983). We also measured several geologically homogeneous areas on LRO NAC images for comparison with Neukum (1983) and areas selected on the LRO WAC mosaic.

The numbers and diameters of the measured craters were exported to Craterstats (Michael and Neukum, 2010; Michael et al., 2012; Michael, 2013; Michael et al., 2016) to derive AMAs using the lunar production and chronology functions of Neukum et al. (2001). We present our CSFD measurement in cumulative plots (Crater Analysis Techniques Working Group, 1979; Neukum et al., 2001; Michael and Neukum, 2010), but also used R-plots (Appendix A6) and differential plots to determine the best fit range. Thus, our determined AMAs can be directly compared with the previously fitted results by Neukum (1983).

# 3. Geological Mapping

Gravity studies of the Mare Tranquillitatis identify the basin as a non-mascon basin, filled with thin lava flows (Staid et al., 1996; De Hon, 2017). De Hon (2017) proposed based on GRAIL gravity anomalies (Zuber et al., 2013) that the irregular shape and topographical differences of Mare Tranquillitatis show that it consists of two overlapping basins.

The lunar module Eagle landed in the southwestern part of Mare Tranquillitatis at 0°41'15" N and 23°26' E, about 44 km northwest of the Moltke crater. We mapped the area surrounding the landing site (southwestern Mare Tranquillitatis) and identified various geological features (Fig.3) including mare units, structural features, crater rays, and different generations of craters.

We used LOLA/Kaguya merged DEM hill shade data as a base map in orthographic projection. The mapping area extends between 15°-33°E and 10°N - 8°S. The digital mapping was completed at a scale of 1:50 K, and the differently sized features used different vertex distances. The smaller geological features have a vertex distance of ~400m, while the larger features have the vertex distance up to 10 km. Identification and confirmation of the geological features required various data sets.

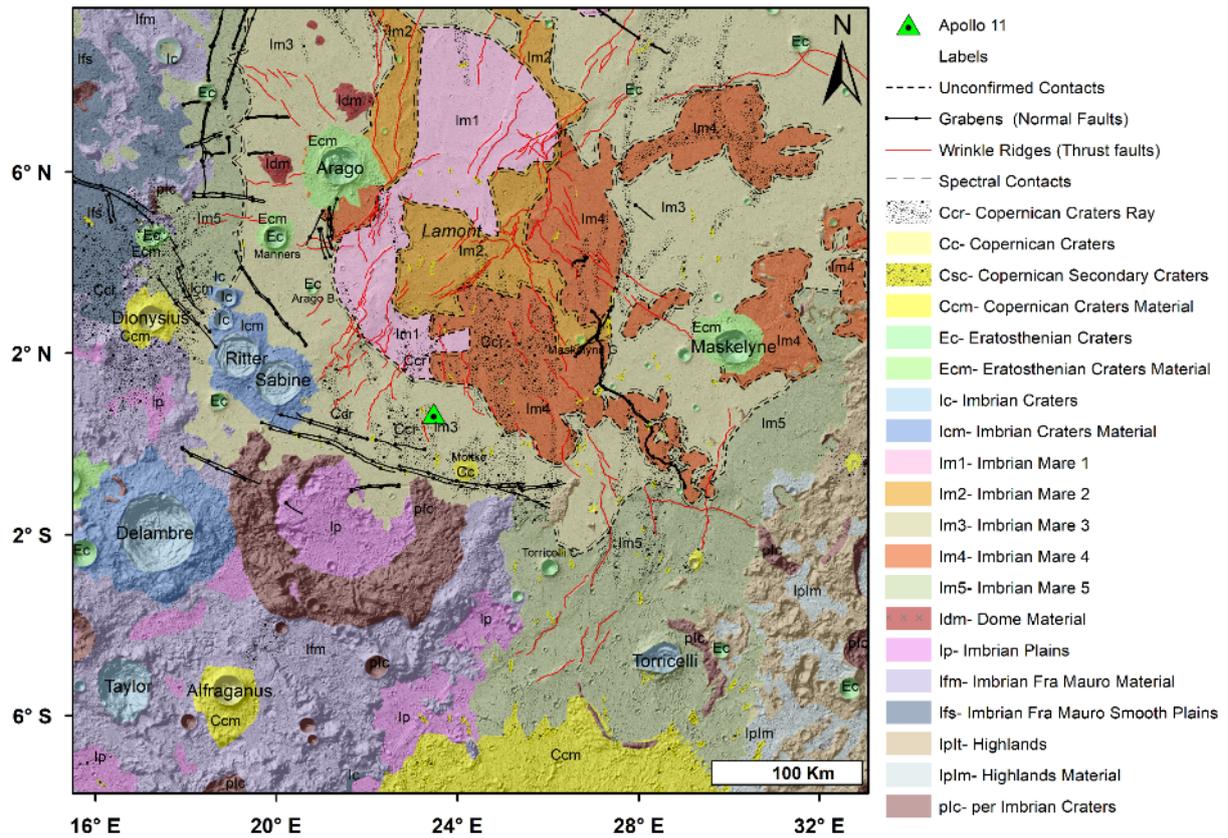

*Fig. 3: The geological map of the southwestern part of Mare Tranquillitatis, showing mare units, the Fra Mauro and Cayley Formations, a radial pattern of wrinkle ridges, and a concentric pattern of rilles, as well as rays and various highland units.*

## 3.1 Mare Basalt Units

The mare basalt units in Mare Tranquillitatis were mapped by earlier workers using different photometric and multispectral data (e.g., Morris and Wilhelms, 1967; Grolier, 1970; Pieters et al., 1994; Staid et al., 1996; Rajmond and Spudis, 2004; Hiesinger et al., 2001, 2003, 2012). In the geologic map of the Sabine D region of the Moon, which contains the Apollo 11 landing site, Grolier (1970) recognized different mare units based on spectral reflectivity during photo-geological studies on the Lunar Orbiter high-resolution photographs: II-H84-2, II-H84-3, II-H85-2, II-H85-3, II-H86-2, II-H86-3, V-H71, V-H72, V-H75, and V-H76. According to Shoemaker et al. (1969) and Grolier (1970), the lunar module landed on the stratigraphically oldest mapped unit. However, the marked gradational boundaries between the mapped units were not definite. In contrast, Staid et al. (1996) proposed that the landing site lies on the youngest and most titanium-rich mare unit in Mare Tranquillitatis, based on a comparison of the radiometric basalt ages (Apollo 11 group A; high-K, high-Ti samples) with the mare units observed with the multispectral Galileo and Clementine data. Most of the previous studies (e.g., DeHon. 1974; Beaty and Albee, 1980; Eugester, 1982; Staid et al., 1996; Rajmond and Spudis, 2004) interpreted the mare units as thin lava flows with an upper Imbrian age, in accordance with the few kipukas of highlands materials in Mare Tranquillitatis.

On the basis of albedo alone, the boundaries of the units are not very well defined. Thus, we used the Clementine UV-VIS color ratio mosaic (Fig. 1b) to identify spectral differences (e.g., Staid et al., 1996; Hiesinger et al., 2000) in the surroundings of the landing site. The Clementine data show clear variations in the concentrations of titanium and iron (Fig. 4, Pieters et al., 1994; Rajmond and Spudis, 2004). However, in Mare Tranquillitatis all basalt units are relatively high in titanium as suggested by the "blueness" of the Clementine false-color ratio data (Fig. 1b, Giguere et al., 2000; Rajmond and Spudis, 2004). Ejecta and rays from various craters caused mixing of low titanium material with the basalts. The scales for the concentrations of the $TiO_2$ and FeO are derived after the sample studies of Giguere et al. (2000) and Clementine-based geological studies of Rajmon and Spudis (2004). Giguere et al. (2000) showed that the $TiO_2$ concentrations of collected basalts samples from the Apollo 11 landing site are higher than 9 wt%. In our new geological map, we defined five mare units representing different lava flows (Fig. 3): Im1, Im2, Im3, Im4, and Im5. We performed CSFD measurements on portions of each flow unit to determine the $N(1)$ values and AMAs for each.

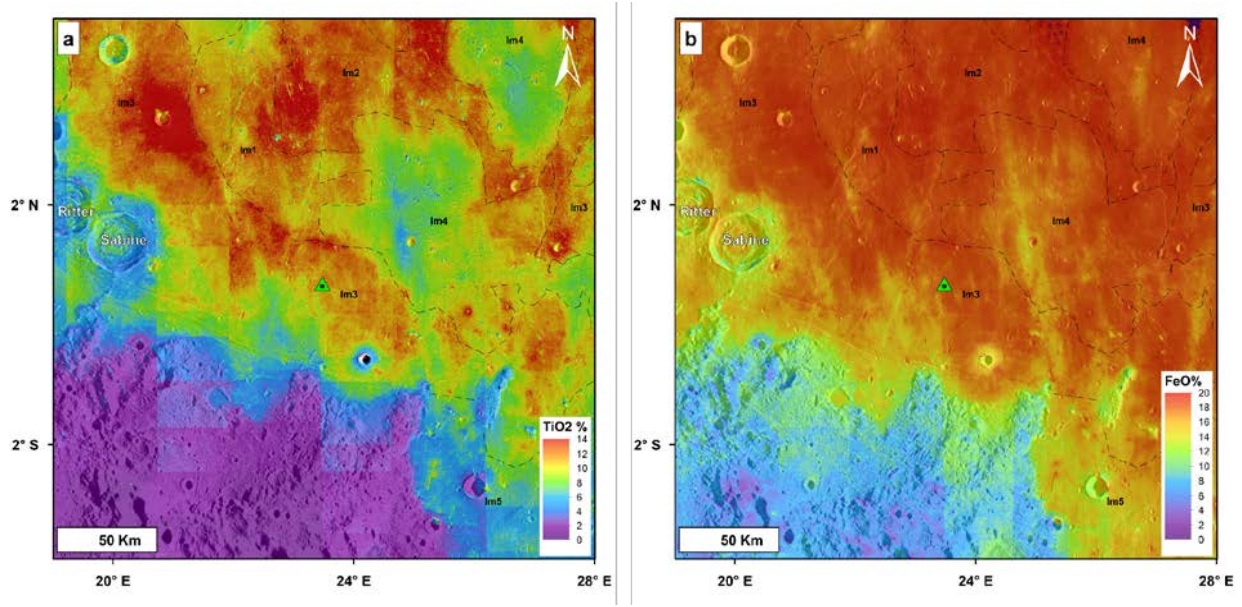

*Fig. 4: Clementine (a) TiO$_2$ and (b) FeO abundance map superposed on LRO WAC mosaic, for the surroundings of the Apollo 11 landing site (green triangle). The spectral boundaries marked as dash lines show the extent of various mare units identified in the mapping area through spectral differences.*

Im1: In the Clementine UV-VIS false-color ratio mosaic, Im1 has an orange color and has superposed yellow-colored craters (Fig. 1b). This unit is well distinguished from the surrounding "blue" mare units. The concentration of TiO$_2$ is about 11 wt% (Fig. 4a) and the concentration of FeO is about 18.5 wt % (Fig. 4b). Lower titanium crater rays overprint the unit (Fig. 4a).

Im1 is surrounded to the east and west by units Im2 and Im4, whereas it is bounded from the north and south by unit Im3 (Fig. 3). The Im1 unit is mapped as a continuous single unit thus, it may be younger than unit Im2.

Im2: The unit Im2 shows a "light blue" color in Clementine false-color mosaic (Fig. 1b), with a strong contrast to the surrounding units. The TiO$_2$ concentration is ~13 wt% (Fig. 4a) and FeO concentration is ~19 wt% (Fig. 4b), which are higher than for unit Im1. The mapped unit extends in north and southeast directions (Fig. 3).

Im3: The Apollo 11 landing site is within this mapped unit. The Clementine data show a "dark blue" color (Fig. 1b). With about 14 wt% and 19 wt% respectively, the TiO$_2$ (Fig. 4a) and FeO (Fig. 4b) concentrations are quite similar to that of unit Im2. This unit surrounds units Im1, Im2 and Im4, making it the largest unit present in our mapped area (Fig. 3). The southern parts of the unit show mixing of low FeO concentrated material transported from the highland units via crater rays.

**Im4:** This unit is characterized by a "bright orange" color in the Clementine false color data (Fig. 1b). The $TiO_2$ concentration is ~9 wt% (Fig. 4a), which is somewhat lower than for the other mare units, perhaps due to the mixing of the low titanium material from crater rays. Whereas, the FeO concentration is ~18 wt% (Fig. 4b). The lobateness of the unit boundary, as well as the differences in albedo and spectral contrasts, suggests that this unit is a distinct mare flow (Fig. 3). Some isolated traces of this unit occur in the eastern part of the study area, indicating superposition of the unit by the younger Im3 lava flows.

**Im5:** This unit exhibits a high albedo in LRO WAC image and blue color with red patches in the Clementine false-color ratio (Fig. 1b). The $TiO_2$ value is 4 wt% (Fig. 4a), while the FeO content is 16 wt % (Fig. 4b) – the lowest of all the studied units. These low values could result from mixing with low titanium and low iron materials from the highlands, because the mapped unit is restricted to the borders of the basin along the foothills of the highlands (Fig. 3).

**Idm - Mare dome material:** In LRO LOLA and SELENE Terrain Camera DEMs, there exist small hills north and northwest of Arago B crater, in mare unit Im3. The gradational boundaries of these topographic highs distinguish them from highlands remnants. Compared to their widths, these dome structures have little topographic relief. They share the same albedo contrast with the surrounding mare material. Similar smaller domes occur in a chain-like fashion and are aligned with a previously identified larger dome, to the north of Arago crater. The domes are not aligned with either wrinkle ridges or graben structures (Fig. 3).

## 3.2 Highland Units

The highlands exhibit clearly different albedos, spectral characteristics, and topography compared to the mare units. They lie in the western and southern parts of the study area. Although the highlands in our study area are characterized by a complex geology, we mapped only the most prominent features in the highland regions, which were relevant for our study of the CSFDs of the geological units in Mare Tranquillitatis. The greater density and larger sizes of craters, in addition to the stratigraphic relationships, indicate that the highlands are older than the mare units.

Morris and Wilhelms (1967) defined the Fra Mauro formation as the ejecta of Imbrian basin in their geological map of the Julius Caesar quadrangle and divided the formation into three units: Fra Mauro hills, Fra Mauro smooth material, and Fra Mauro buried material. Our map mostly agrees with the Morris and Wilhelm (1967) distribution and interpretation of the units, but not with the extents of these units (Fig. 3). Foretzzo and Hare (2013) digitally renovated map of 1:5M did not clearly distinguish the plain material of Fra Mauro Formation from other Imbrian plain units.

**Ifm - Fra Mauro hills:** The highlands to the west of the landing site mainly consist of Imbrium ejecta classified as the Fra Mauro Formation (Fig. 5a), which is characterized by a northwest-southeast lineated terrain, oriented radially to the Imbrium basin.

**Ifs - Fra Mauro Smooth terrain:** In the northwest region, we mapped the smooth plains as a sub-unit of the Fra Mauro Formation. This unit is less hummocky than the Fra Mauro hills (Ifm), but shows a fine lineation similar to Fra Mauro formation (Fig. 5b). This unit might consist of mare material mixed with highland material.

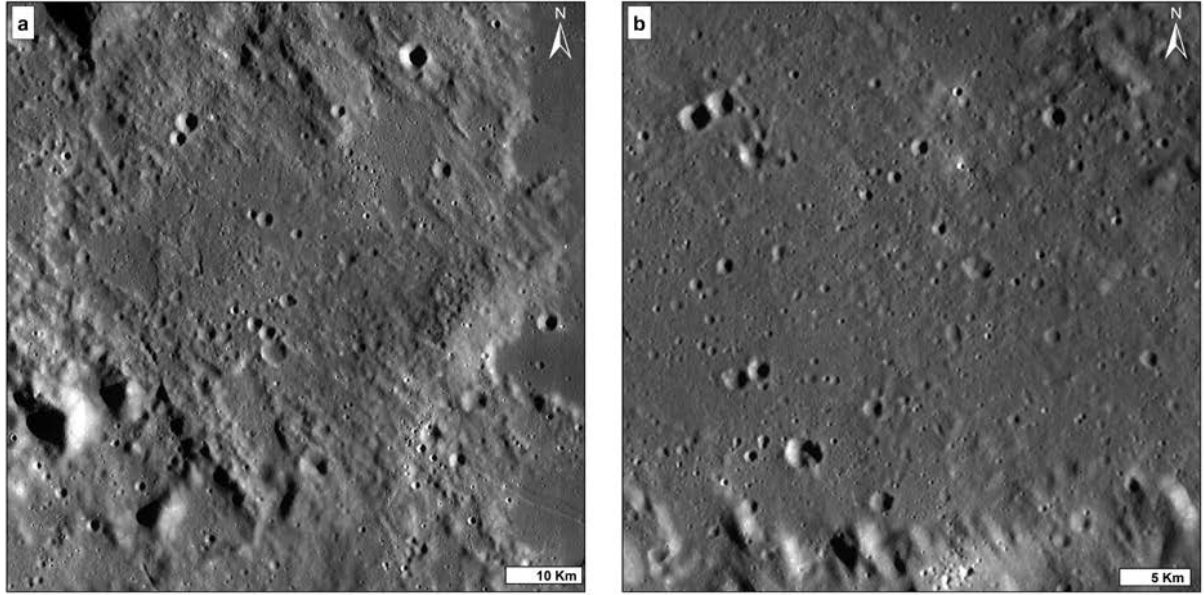

*Fig. 5: Two units of the Fra Mauro Formation were mapped in the northwest of the landing site in the LRO WAC mosaic (reference map A2 in Appendix): (a) The Ifm unit shows lineations in a northwest-southeast direction, and (b) the Ifs unit, which is smoother than Ifm, shows only faint lineations similar to the Ifm unit.*

**Ip – Imbrian Plains:** Morris and Wilhelms (1967) named this unit the Cayley Formation (Ica), which is quite similar to the smooth areas of the Fra Mauro formation, except for being more hummocky (Fig. 6a). Mapped as patches in the highlands, this unit probably represents accumulated volcanic ash material or ballistically deposited ejecta layers from surrounding craters (Morris and Wilhelms, 1967). Foretzzo and Hare (2013) designated most of the plain material in the highlands as Ip. Our map distinguishes Ip from the Ifs and IpIm units on the basis of albedo, spectral, and stratigraphical differences. This unit was also mapped as lunar light plains by Meyer and Boyd (2018).

**IpIt - Highland (terrain):** This unit includes all highland terrain materials that are not further subdivided and may consist of more than one geological unit. The unit likely includes the ejecta from pre-Imbrian or Imbrian basins. Although, in the mapping area the origin of this unit is unclear (Fig. 6b).

**IpIm - Highland plains material:** IpIm includes all highland plains materials that cannot be classified as Ifs or Ip terrains. This unit is smooth; however, it is the most hummocky of the three plains units (Ifs, Ip and IpIm) (Fig. 6b). Wilhelms (1972) compared this unit to Ip but considered it as intercrater materials of pre-Imbrian age, whereas Foretzzo and Hare (2013) mapped it as Ip. In our interpretation, it may have the same origin as the Cayley formation with a different thickness, with accumulated ejecta from the Copernican-aged Theophilus crater and/or other young craters.

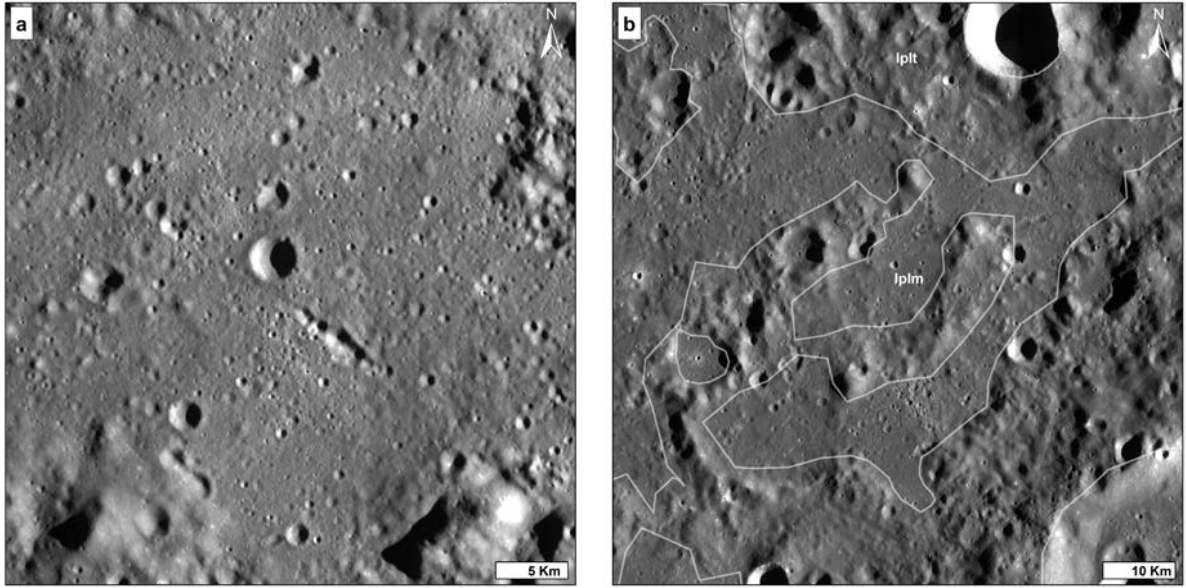

*Fig. 6: Additional highland units (reference map A2 in Appendix) identified in the LRO WAC mosaic include: (a) Ip unit, which is also known as the Cayley Formation (Morris and Wilhelms, 1967), and (b) IpIt and IpIm units, which may contain more than one unit of Imbrian and pre-Imbrian ages.*

## 3.3 Structures

The newly derived geologic map shows a radial pattern of thrust faults recognized as wrinkle ridges (Fig 7a.) in the southwestern part of Mare Tranquillitatis. In the middle of the map is an unusual circular ridge pattern called the Lamont ridges (Morris and Wilhelms, 1967; Shoemaker et al., 1969). Shoemaker et al. (1969) interpreted the Lamont ridges as localized wrinkle ridges over a shallow mare-covered crater. However, we did not observe other evidence for a buried crater in our data sets.

Graben structures or arcuate rilles (Fig 7b.) form a semi-concentric radial pattern around the mare units, which follow the southwestern edge of Mare Tranquillitatis. However, in our study area a few straight graben could also be mapped in the highlands (Fig. 3). About 140 km west of the landing site, there is a sinuous rille oriented from northwest to southeast. This rille can be interpreted as an extinct lava tube associated with the mare unit Im4, as it follows the same direction as the mare flow. Though the sinuous rille is traceable in parts, no prominent and associated pits or skylight craters were found.

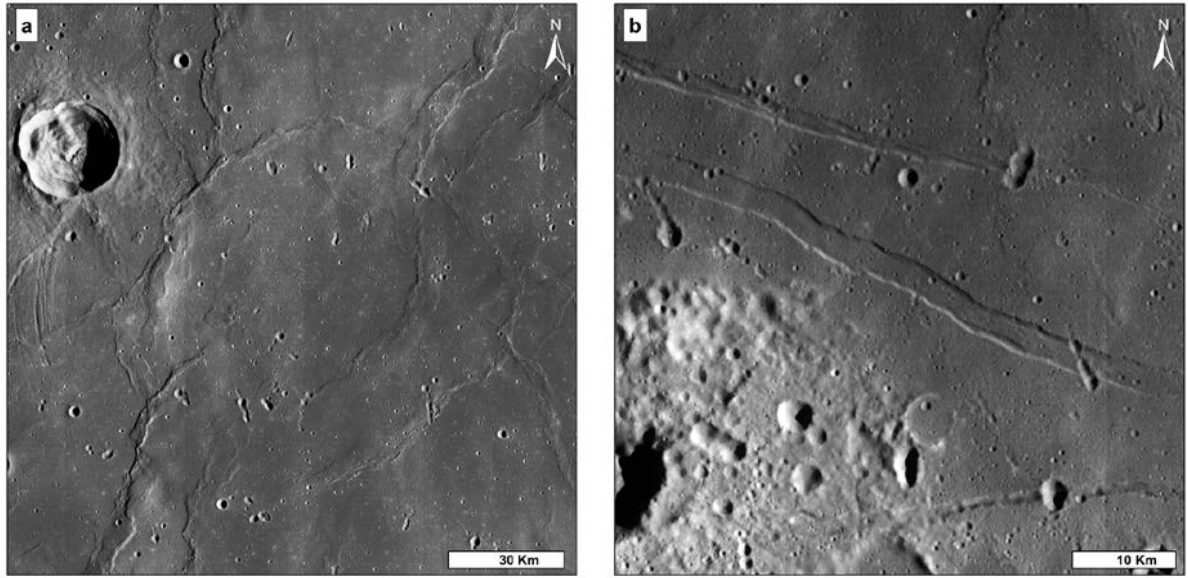

*Fig. 7: Identified structures (reference map A3 in Appendix) in the LRO WAC mosaic include: (a) wrinkle ridges forming radial pattern and the Lamont ring structure in the southwestern Mare Tranquilities, and (b) graben or rilles, along the edge of the mare units and the radial pattern of the wrinkle ridges.*

The elevation profile and stratigraphic relationships of the mapped rilles and ridges in the mare area show they are younger than the youngest mare units. Golombek and McGill (1983) proposed that the wrinkle ridges and rilles are formed by stresses generated during subsidence of the basin.

## 3.4 Craters

At the landing site, crater sizes range up to several tens of meters in diameter, with some craters that are interpreted as secondaries from Theophilus (Morris and Wilhelms, 1967; Aldrin et al., 1969; Shoemaker et al., 1969; Grolier, 1970). West crater is situated approximately 400 m east of the landing site, has a sharp rim, and a very blocky surface. The 12 m long, 6 m wide, 1 m deep Double crater is located to the southwest of the lunar module (Fig. 2). About 50 m east of the lunar module is Little West crater (Fig. 2), 33 m across and 4 m deep (Aldrin et al., 1969; Shoemaker et al., 1969; Grolier, 1979, Beaty and Albee, 1980).

The crater population around the landing site is classified into different age classes (e.g., Morris and Wilhelms, 1967; Wilhelms, 1972, 1987; Grolier, 1970; Fortezzo and Hare, 2013) such as pIc- pre-Imbrian craters, Ic- Imbrian craters, Ec- Eratosthenian Craters, and Cc- Copernican craters

*pIc:* The walls of craters in the highlands are commonly highly degraded; the craters themselves are partially filled with mare units and/or other craters material. They have no clearly defined ejecta in LRO and Clementine data. These craters are mostly covered by Imbrium ejecta material (Fig. 8a.), meaning they are pre-Imbrian in age.

*Ic and Icm:* Imbrian craters (Ic) are old craters that are often filled by mare units. The rims of these craters are slightly degraded, but still visible (Fig. 8.b). The ejecta (Icm) of these craters is still somewhat visible, but is moderately degraded. The extent of the ejecta can be traced with spectral data. These craters are either Imbrian in age or late Imbrian in age, as they completely overlie the Imbrium ejecta material.

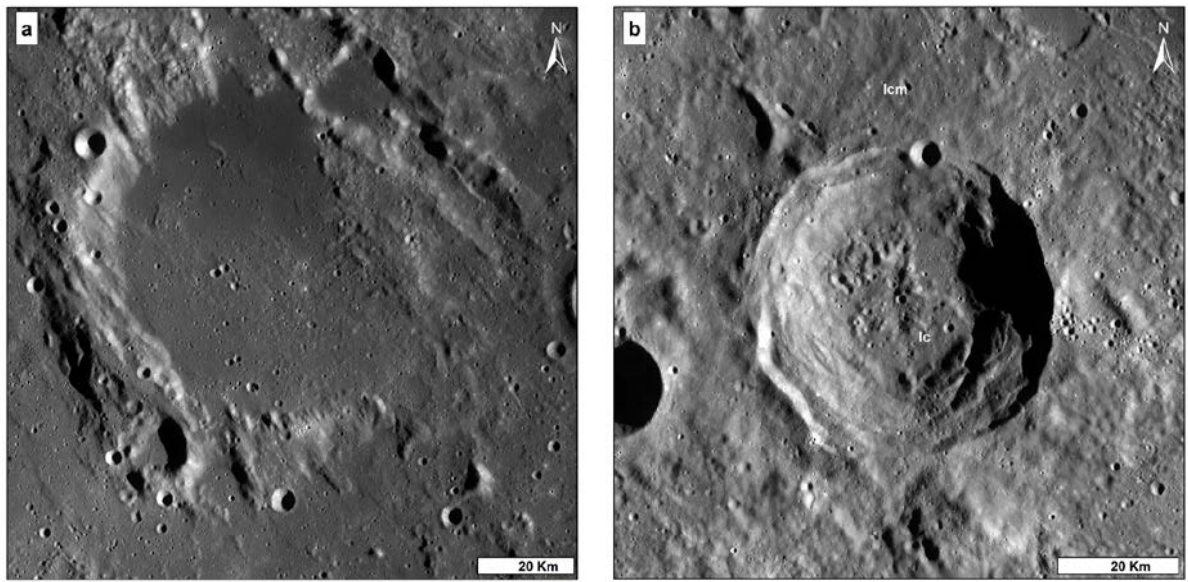

*Fig. 8: Craters in the LRO WAC mosaic (reference map A4 in Appendix) include: (a) pIc, highly degraded pre-Imbrian Julius Caeser crater and (b) Ic, Imbrian craters with visible, slightly degraded rims and partially degraded ejecta (Icm).*

*Ec and Ecm:* Eratosthenian craters exhibit sharp rims. Like Imbrian craters, they also show evidence of partially degraded ejecta, which can be mapped with spectral data. Ec is distinguished from Ic on the basis of their stratigraphic position and crater rim's degree of the degradation (Fig 9a.). We used Clementine data to delineate the extent of the ejecta deposits. Small Erathosthenian craters do not show any clear traces of ejecta in these data, thus separating them from younger Copernican craters.

*Cc and Ccm:* The young Copernican craters were classified on the basis of sharp rims and obvious ejecta deposits, which can be easily traced in both albedo and spectral data. These craters show bright ejecta material (Ccm) and rays extending away from them (Fig 9b.). For example, Moltke is a Copernican crater, which contributed ray material to the Apollo 11 landing site.

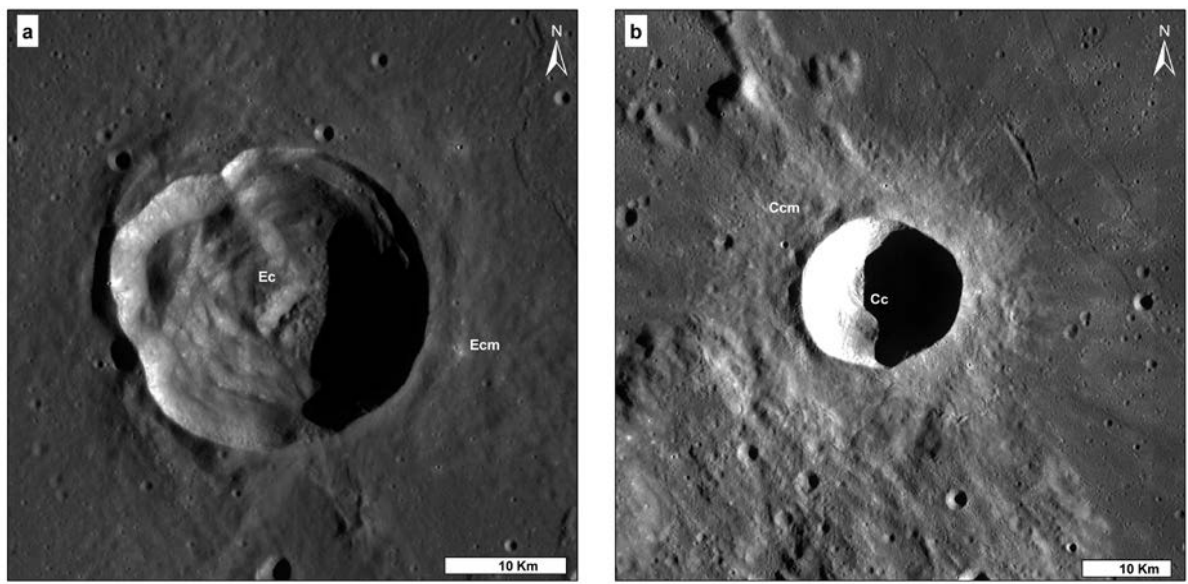

*Fig. 9: Craters identified in the LRO WAC mosaic (reference map A4 in Appendix) include: (a) Ec, Eratosthenian crater with partially degraded rim and ejecta Ecm, and (b) Cc, Copernican crater with a well-defined rim, ejecta (Ccm), and rays (Ccr).*

## 3.5 Rays (Ccr)

The area surrounding the landing site is characterized by discontinuous ejecta (rays or Ccr unit in Fig. 3) from nearby and distant craters (Morris and Wilhelms, 1967; Aldrin et al., 1969; Shoemaker et al., 1969). In our map, these rays are shown as a separate geological unit on the basis of spectral contrast and the presence of the secondary craters. Most of the rays and secondary craters in the map area can be traced to Theophilus (Fig. 3 shows the ejecta of Theophilus south of Torricelli crater). Some rays with a northeast-southwest orientation may have been emplaced by Tycho, Alfraganus, or other younger craters (Morris and Wilhelms, 1967; Shoemaker et al., 1969). Approximately 15 km west of the landing site is a ray that trends north-northeast. Shoemaker et al. (1969) suggested this ray was formed by either Alfraganus or Tycho (Fig. 3).

# 4. Crater Size-Frequency Distributions (CSFDs) Measurements

To test and improve the lunar chronology, detailed CSFD measurements were done on LRO WAC and NAC images (Fig. 10, 11). The area used to determine the CSFD measurements were newly selected around the landing site while avoiding contamination from secondary crater material and other geological units (Fig. 3). Although, the effect of count area size on the accuracy of AMAs is still being investigate (e.g, Hiesinger et al. 2012, Pasckert et al. 2015; van der Bogert et al.,2015), we selected reasonable area sizes and crater diameter ranges for the determination of $N(1)$ values for avoiding large uncertainties (see Tables, 1, 2 and 3). The crater diameters measured for the age determinations range from 10 m to 3 km. For the comparison of the calibration points (Table 4), we also measured CSFDs for the area previously defined by Neukum (1983), using the LRO WAC mosaic. For the confirmation of our results, we compared our studies with other studies mentioned in Table 4.

## 4.1 LRO WAC measurements

We selected two areas on the LRO WAC mosaic for comparative CSFD measurements. These areas represent two geologically homogeneous units: A mare unit named WAC_Im3 and a ray unit, WAC_Ccr (Fig. 10a). We selected an area on the ray material in order to investigate whether there was any effect of the ray on the CSFD. However, both areas give similar $N(1)$ values of $7.13 \times 10^{-3} \pm 1.19$ km$^2$ and $7.03 \times 10^{-3} \pm 1.17$ km$^2$, respectively (Fig. 10a'). As such, we merged the two areas to improve the statistics of the result, and get an $N(1)$ value of $6.88 \times 10^{-3} \pm 0.611$ km$^{-2}$. Our recount of the Neukum (1983) area, fit with the Neukum et al. (2001) production function gives an $N(1)$ value of $6.47 \times 10^{-3} \pm 0.496$ km$^{-2}$. The summary our results gained from LRO WAC data is shown Table. 1.

Table 1. Areas, fit ranges, N(1)'s and model ages derived from the LRO WAC data.

| Area Name | Area [km$^2$] | No. of craters fit | Fit range (km) | $N_{cum}$(D≥1 km) [km$^{-2}$] | $N_{cum}$(D≥1km) [km$^{-2}$] error | AMA [Ga] | AMA [Ga] error |
|---|---|---|---|---|---|---|---|
| WAC_Im3 | 444 | 36 | 0.4-1.2 | 7.13x10$^{-3}$ | ±1.19x10$^{-3}$ | 3.61 | ±0.027 |
| WAC_Ccr | 356 | 36 | 0.45-1.5 | 7.01x10$^{-3}$ | ±1.17x10$^{-3}$ | 3.61 | ±0.04 |
| WAC_merge | 8 | 127 | 0.4-1.5 | 6.88x10$^{-3}$ | ±6.11x10$^{-4}$ | 3.61 | ±0.022 |
| A11_WAC (Neukum, 1983) | 1590 | 170 | 0.45-2.5 | 6.47x10$^{-3}$ | ±4.96x10$^{-4}$ | 3.59 | ±0.02 |

## 4.2 LRO NAC measurements

We selected four geologically homogenous areas in NAC frames M102000149 and M162154734: (a) Im3_LS, (b) Im3_LS_North, (c) Im3_LS_South, and (d) Ccr_LS (Fig. 11.a, a'). The area Im3_LS is a homogeneous area surrounding the lunar module. Extra vehicular activities (EVAs) during the Apollo 11 mission took place within the boundaries of this area. The calculated $N(1)$ of this area is 5.74x10$^{-3}$±0.99 km$^{-2}$. Area Im3_LS_South, to the south of the landing site, yields an $N(1)$ of 3.95x10$^{-3}$±0.69 km$^{-2}$. Area Im3_LS_North, to the north of the landing site, gives $N(1)$ of 6.97x10$^{-3}$±0.76 km$^{-2}$. Area Ccr_LS was measured on an area containing a ray to investigate whether the CSFD here was affected by the emplacement of the ray, and yields an $N(1)$ of 6.97x10$^{-3}$ ±1.29 km$^{-2}$, which is similar to $N(1)$ value of Im3_LS_North but with higher uncertainty. The randomness analysis (Michael et al., 2012) of the count areas shows the presence of potential secondary clustering in diameter ranges below 50 m. Thus these diameter ranges were not used for fitting the $N(1)$ values. If the CSFD measurements of these areas are merged (Fig. 11b, b'), they result in an overall an $N(1)$ of 6.42x10$^{-3}$±0.54 km$^{-2}$. The comparison and details of the measured values are summarized in Table 2.

Table 2. Areas, fit ranges, N(1)'s and model ages derived from the LRO NAC data.

| Area Name | Area [km$^2$] | No. of craters fit | Fit range (m) | N$_{cum}$(D≥1km) [km$^{-2}$] | N$_{cum}$(D≥1km) [km$^{-2}$] error | AMA [Ga] | AMA [Ga] error |
|---|---|---|---|---|---|---|---|
| Im3_LS | 15.6 | 33 | 220-500 | 5.74x10$^{-3}$ | ±9.99x10$^{-4}$ | 3.56 | ±0.05 |
| Im3_LS_South | 13.6 | 33 | 170-500 | 3.95x10$^{-3}$ | ±6.88x10$^{-4}$ | 3.42 | ±0.1 |
| Im3_LS_North | 76.2 | 84 | 250-900 | 6.97x10$^{-3}$ | ±7.6x10$^{-4}$ | 3.61 | ±0.022 |
| Ccr_Ls | 29.8 | 29 | 250-700 | 6.97x10$^{-3}$ | ±1.29x10$^{-3}$ | 3.61 | ±0.04 |
| NAC_merge | 135 | 140 | 230-900 | 6.42x10$^{-3}$ | ±5.43x10$^{-3}$ | 3.59 | ±0.022 |

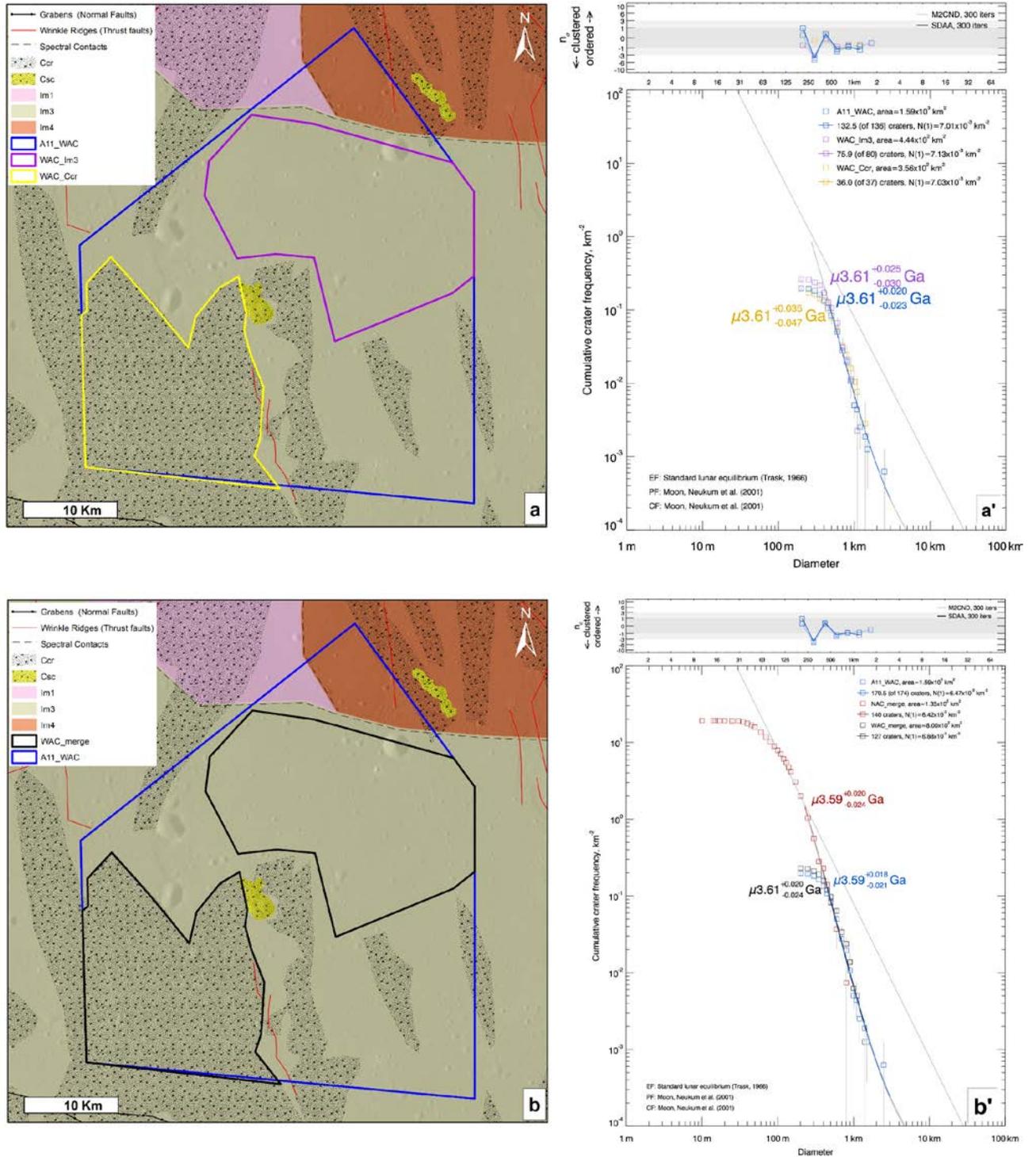

*Fig. 10 (a) Two geological units were measured in LROC WAC mosaic, i.e., a mare basalt (purple) and a ray (yellow) for comparison with the Neukum (1983) count area (blue). (a') The N(1) valued obtained on the cumulative plot for remeasured Im3, Ccr and Neukum (1983) areas (b) New areas measured on LRO WAC data (black) and Neukum (1983) count area (blue). (b') The merged WAC-derived (black) AMA value is identical to that derived from the WAC CSFD measurements (blue).*

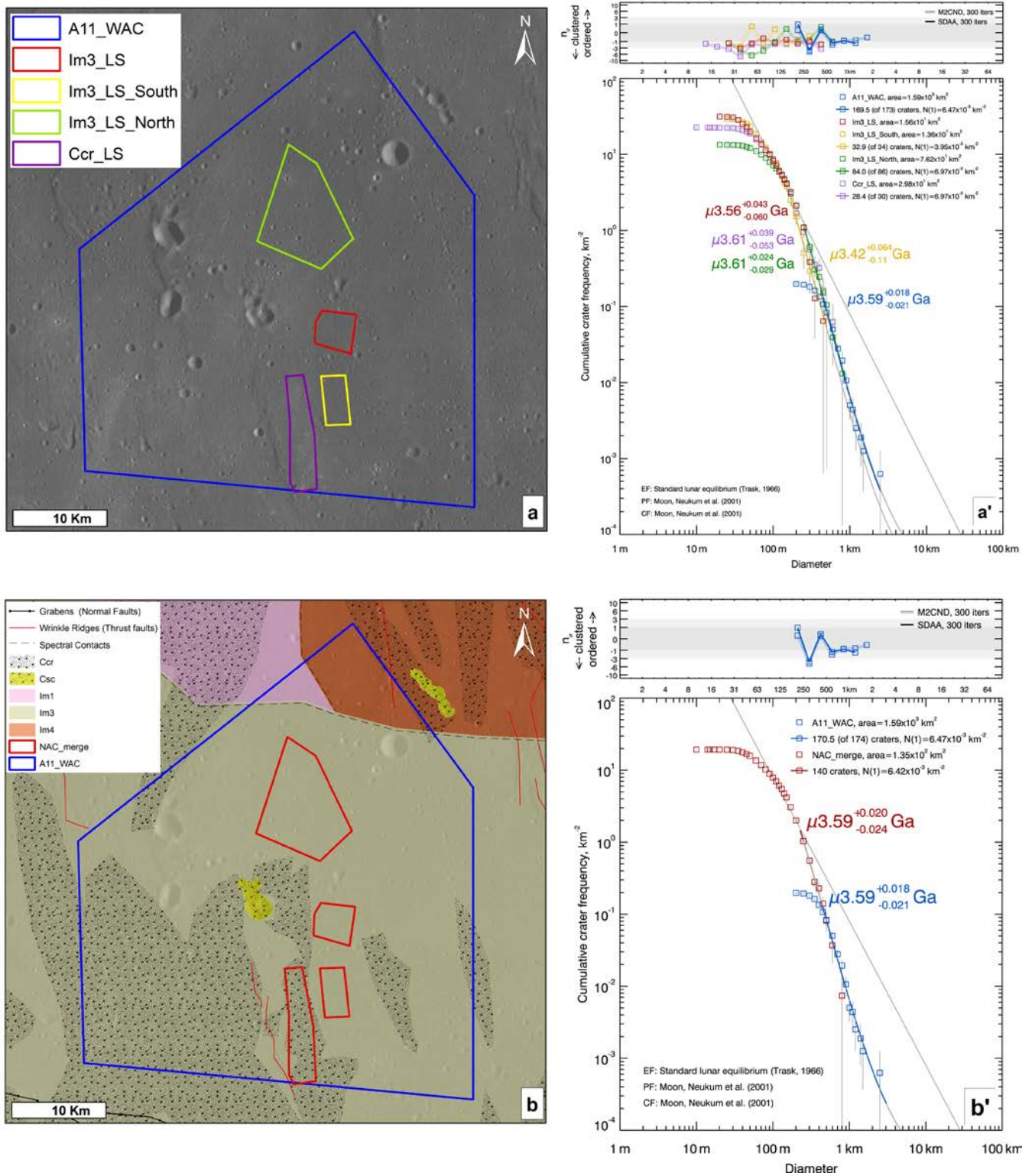

*Fig. 11: CSFD measurement areas, cumulative frequency plots, and absolute model age fits. (a) The original area mapped by Neukum (1983) (blue) was remeasured on the LROC WAC mosaic, in addition to the small areas marked in red (*Im3_LS, *landing site), yellow (*Im3_LS_South*), green (*Im3_LS_North*) and violet (*Ccr_LS*), which were measured on LRO NAC data. (a') The CSFD from the Neukum (1983) area is compared with the areas measured on NAC data. (b) Areas measured on LRO WAC data (blue)*

*and LRO NAC data (red). (b') The merged NAC-derived (red) AMA and N(1) value, is identical to that derived from the WAC CSFD measurement (blue).*

# 5. Discussion

## 5.1 Discussion of Various Geological Units

We found that the basalt units Im1, Im2, Im3, and Im4 may have differences in age of a few hundred million years, as suggested by our CSFD measurements (Table. 3). The CSFD of the unit Im1 allows us to fit two $N(1)$ values $5.26 \times 10^{-3} \pm 0.445$ km$^{-2}$ and $1.52 \times 10^{-2} \pm 0.128$ km$^{-2}$. The CSFD of unit Im2 yields $N(1)$ value of $8.71 \times 10^{-3} \pm 0.736$ km$^{-2}$. Unit Im3 has $6.90 \times 10^{-3} \pm 0.529$ km$^{-2}$. The $N(1)$ value for Im4 is $8.25 \times 10^{-3} \pm 0.697$ km$^{-2}$. On the basis of our CSFD measurements, we find two $N(1)$ values of $9.73 \times 10^{-3} \pm 0.822$ km$^{-2}$ and $3.88 \times 10^{-2} \pm 0.3286$ km$^{-2}$ for unit Im5. Some of the basalts (e.g., Im3, Im4) were modified by the formation of wrinkle ridges and the emplacement of ray materials. Thus, the CSFDS of the small craters measured in the NAC images for these areas might have been affected by these processes. However, CSFD measurements on the WAC data, show no effects from the ray material on the resulting AMAs around the landing site, suggesting that the ray materials are rather thin and did not affect the size range of craters we used for determining the $N(1)$ values and absolute model ages.

*Table 3. Compared crater retention values and model ages of different mare units.*

| Unit | Area [km$^2$] | No. of craters fit | Fit range (km) | $N_{cum}(D \geq 1$ km) [km$^{-2}$] | $N_{cum}(D \geq 1$ km) [km$^{-2}$] error | AMA [Ga] | AMA [Ga] error |
|---|---|---|---|---|---|---|---|
| Im1 | 669 | 58 | 0.45-1.4 | $5.26 \times 10^{-3}$ | $\pm 4.45 \times 10^{-4}$ | 3.53 | $\pm 0.04$ |
| | | 3 | 1.1-2.5 | $1.52 \times 10^{-2}$ | $\pm 1.28 \times 10^{-3}$ | 3.77 | $\pm 0.13$ |
| Im2 | 750 | 71 | 0.491-2 | $8.71 \times 10^{-3}$ | $\pm 7.36 \times 10^{-4}$ | 3.66 | $\pm 0.025$ |
| Im3 | 1430 | 121 | 0.45-3 | $6.90 \times 10^{-3}$ | $\pm 5.29 \times 10^{-4}$ | 3.61 | $\pm 0.022$ |
| Im4 | 499 | 114 | 0.4-1.1 | $8.25 \times 10^{-3}$ | $\pm 6.97 \times 10^{-4}$ | 3.65 | $\pm 0.02$ |
| Im5 | 645 | 39 | 0.6-1.2 | $9.73 \times 10^{-3}$ | $\pm 8.22 \times 10^{-4}$ | 3.68 | $\pm 0.03$ |
| | | 3 | 2-3 | $3.88 \times 10^{-2}$ | $\pm 3.28 \times 10^{-3}$ | 3.93 | $\pm 0.01$ |

We compared the new mare unit ages and boundaries with those of Hiesinger et al. (2000). Our study area contains the following units of Hiesinger et al. (2000): T1, T3, T6, T9, T11, T12, T14, T15, T17, T18, T20, T21, T22, T23, T26, and T27 (Fig. 12. b). These mare units were mapped on the basis of Galileo spectral information, thus, showing a few differences compared with our study, which uses Clementine data. Staid et al. (1996) explained that although Clementine data used coinciding filters with Galileo data, due to the higher resolution (pixel scale of 150-300 m) of the Clementine data compared to the Galileo data (pixel scale of 1.5-2 km), the spectral boundaries of the various basalts units appear clearer. The Hiesinger et al. (2000) mapping was mainly focused on the spectral differences in the lunar maria, whereas our detailed mapping identified various other geological units. We also used $TiO_2$ and FeO data to confirm our mare unit boundaries (Fig. 4). Unit T17 defined by Hiesinger et al (2000) covers an area around the landing site and yields $N(1)$ $7.60 \times 10^{-3} \pm 1.00$ $km^{-2}$, shows a model age of 3.61 Ga, which compares to the newly mapped unit Im3, which has an $N(1)$ value of $6.42 \times 10^{-3} \pm 0.54$ $km^{-2}$ and AMA of $3.61 \pm 0.023$ (see Table 3) and also covers an area around the landing site. We show individual cumulative plots for the CSFD measurements of these units in Appendix A5.

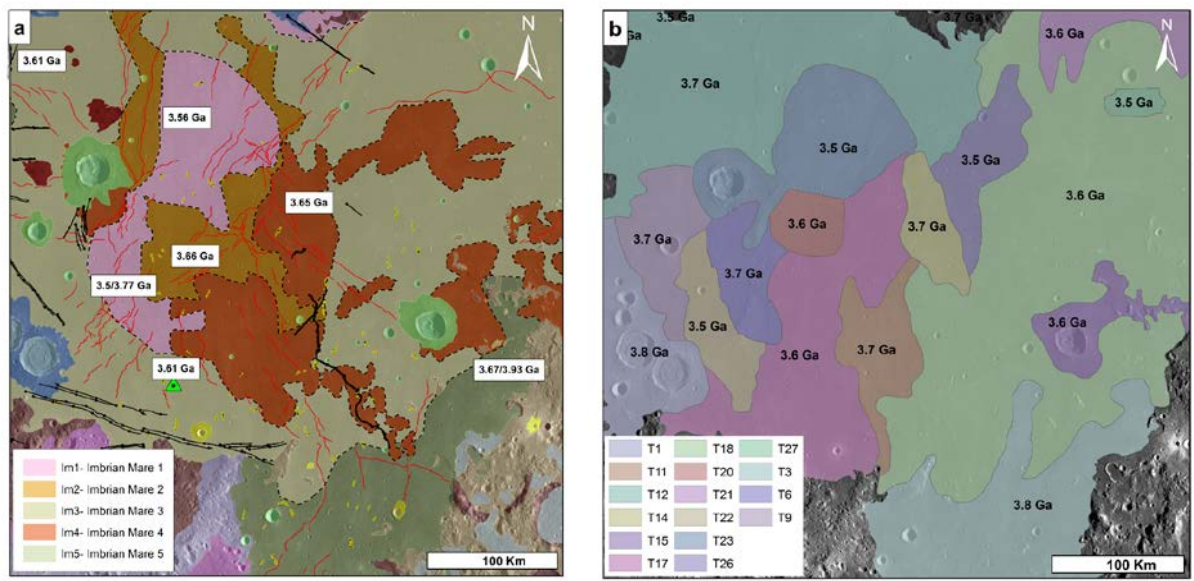

*Fig. 12: (a) The geological map showing the ages of different mare units identified in the southwestern part of Mare Tranquillitatis around the Apollo 11 landing site (green triangle). (b) A map of the mare boundaries defined by Hiesinger et al. (2000) using Galileo*

*multispectral data. Our new ages are consistent with the ages determined by Hiesinger et al. (2000), despite the units having somewhat different boundaries.*

## 5.1 Discussion of CSFD measurements

Neukum (1983) measured two types of lava flows around the lunar module, both showing distinct $N(1)$ values. In his work, the age of the old flow (3.72 ±0.1 Ga) has an $N(1)$ value of $9.0 \times 10^{-3}$ km$^{-2}$, while the young flow (3.53 ±0.05 Ga) has an $N(1)$ of $6.4 \times 10^{-3}$ km$^{-2}$ (Neukum, 1983). This work was followed by Stöffler et al. (2001, 2006), who reassessed the radiometric ages assigned to the "young" and "old" flows based on sample ages. Robbins (2014) interpreted his new $N(1)$ $7.67 \times 10^{-3}$ ±1.02 km$^{-2}$ from a larger area surrounding the landing site to be closer to the "old" mare units. Our re-measurement of the Neukum (1983) area found an $N(1)$ of $6.42 \times 10^{-3}$ km$^{-2}$ with NAC data, and an $N(1)$ of $6.11 \times 10^{-3}$ km$^{-2}$ with WAC data. Our results are consistent with the AMA of the young lava flow of Neukum (1983), but is inconsistent with the old lava flows AMA.

*Table 4. Comparison of crater retention values and model ages of different mare units from different investigators.*

|  | Unit | Area [km$^2$] | Number of craters fit | $N_{cum}(D\geq1$km$)$ [km$^{-2}$] | $N_{cum}(D\geq1$km$)$ [km$^{-2}$] error | AMA [Ga] | AMA [Ga] error |
|---|---|---|---|---|---|---|---|
| This work | NAC_merge (Im3) | 135 | 140 | $6.42 \times 10^{-3}$ | ±$0.543 \times 10^{-3}$ | 3.59 | ±0.022 |
|  | WAC_merge (Im3) | 800 | 127 | $6.88 \times 10^{-3}$ | ±$0.615 \times 10^{-3}$ | 3.61 | ±0.022 |
| Robbins, 2014 | Apollo 11 | 12,828 | 604 | $7.67 \times 10^{-3}$ | ±$1.02 \times 10^{-3}$ | 3.72 |  |
| Marchi, 2009 [NEO] | MT (young) |  |  | $9.30 \times 10^{-3}$ |  | 3.58 |  |
|  | MT (old) |  |  | $1.836 \times 10^{-3}$ |  | 3.80 |  |
| Hiesinger et al., 2000 | T17 | 3754 |  | $7.60 \times 10^{-3}$ | ±$1.00 \times 10^{-3}$ | 3.63 | ±0.07 |

| Neukum, 1983 | A11 "young" | ~1590 | | $6.40 \times 10^{-3}$ | $\pm 2.00 \times 10^{-3}$ | 3.53 | ±0.05 |
| | A11 "old" | ~1590 | | $9.00 \times 10^{-3}$ | $\pm 1.80 \times 10^{-3}$ | 3.72 | ±0.1 |

The area defined by Neukum (1983) for CSFD measurements crosses three different mare units as defined in our geological map (Fig. 10, 11). This area yields an AMA of about 3.59 Ga on our WAC CSFD measurements. However, neither in our map nor in the CSFDs did we observe any evidence for an old flow exposed at the surface near the landing site area. The $N(1)$ value $6.90 \times 10^{-3} \pm 0.529$ km$^{-2}$ of the mare unit Im3 is also consistent within error with the $N(1)$ value $6.42 \times 10^{-3} \pm 0.54$ km$^{-2}$ measured around the landing site with NAC data and $6.88 \times 10^{-3} \pm 0.611$ km$^{-2}$ measured on the WAC data. However, the mare unit Im5 is the oldest mare unit mapped at the edges of the basin has $N(1)$ of $3.88 \times 10^{-2} \pm 0.3286$ km$^{-2}$. Either this unit is missing around the landing site or it is buried underneath the younger units at the landing site, and is thus not visible in the data we used for mapping and CSFD measurements.

We prefer the $N(1)$ value derived on NAC data as the one to represent the geological unit from which the youngest basalt samples come, in part because these carefully chosen areas for tend to be less contaminated by secondary craters identified by randomness analyses. Although we paid close attention to avoid any obvious secondary clusters and chains in our CSFD measurements, the randomness analyses (Michael et al., 2012) mostly indicate that craters smaller than ~50m show some degree of clustering. However, these crater diameters were not used to fit the $N(1)$'s or AMAs and thus do not affect our results.

# 6. Sample Studies

## 6.1 Sample Collection

During the Apollo 11 mission, 22 kg of samples were collected in the vicinity of the lunar module (Aldrin et al., 1969; Kramer et al., 1977; Meyer, 2012). The landing area is covered with porous and weakly coherent regolith. Hess and Calio (1969) described the landing site as consisting of debris ranging from 1 meter down to microscopic particles. Initially, one kilogram of contingency sample was collected adjacent to the lunar module (Fig. 13). Later, another fifteen kilograms of samples were collected as scooped rock and soil, in addition to two cores. The samples in bulk were collected as "documented samples" from an area 10 to 15 m southeast of the lunar module, shown in Fig. 14 (Aldrin et al., 1969; Kramer et al., 1977; Meyer, 2012). The regolith is estimated to be approximately 5 m thick in the landing region (Hess and Calio, 1969). According to the hypothesis of Beaty and Albee (1980), the samples may have been excavated by the formation of a single nearby crater such as West crater. Aldrin et al., (1969) suggested that West crater may have excavated material from a depth of 30 m, and could have provided direct samples of the bedrock. According to this theory, the regolith is derived from local bedrock (Aldrin et al., 1969; Beaty and Albee, 1980).

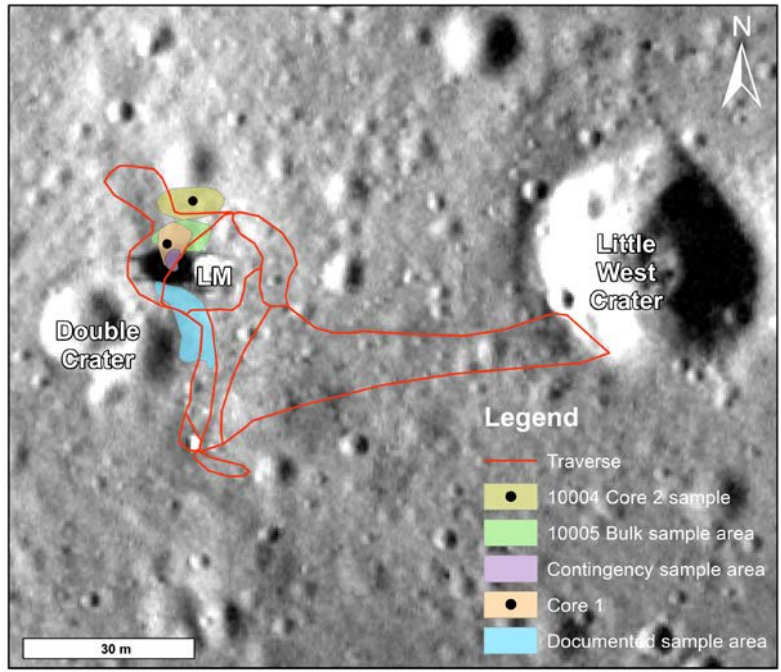

*Fig. 13: Mapped astronauts traverse around the Apollo 11 landing module and sample collection areas. Samples were collected as cores, bulk material, and scooped soils as defined by Kramer et al. (1977), on basis of the Preliminary Science Report Aldrin et al. (1969) and astronauts accounts.*

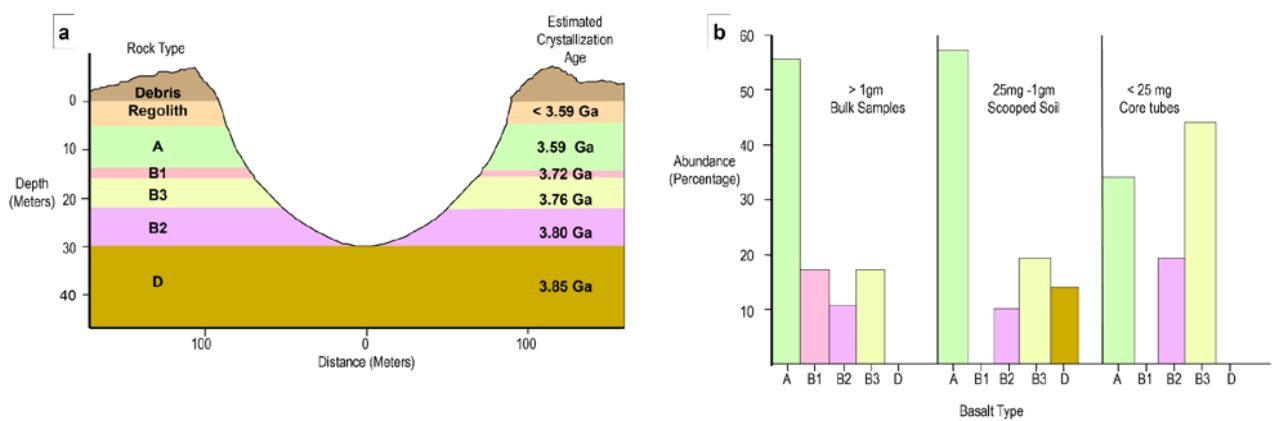

*Fig. 14: (a) The abundance of five types of basalts (A, B1, B2, B3, and D) in the collected bulk, scooped soil, and core samples. (b) Hypothetical sketch of West crater showing the sequence of basalts interpreted to form the determined radiometric ages. Modified after Beaty and Albee (1980) using ages of the units as defined by Stöffler et al. (2001, 2006).*

## 6.2 Radiometric Analysis of Basalts

Petrological and radiometric studies divided the basalt samples into five different groups: group A, B1, B2, B3, and D (Jerde et al., 1994). Group A consists of high-K, high-REE (rare earth element) basalts; group B (including B1, B2, B3) are low-K, low-REE basalts; and group D are low-K, high-REE basalts (Jerde et al., 1994; Stöffler et al., 2006). The five groups of basalts are interpreted to represent different lava flows (Beaty and Albee, 1980; Jerde et al., 1994; Stöffler et al., 2006). The exposure ages indicate that the high potassium basalts were located on the surface, whereas the low potassium basalts do not have outcrops near the landing site (Aldrin et al., 1969; Beaty and Albee, 1980; Stöffler et al., 2006).

The crystallization ages determined via $Ar^{39}$-$Ar^{40}$ (Guggisberg et al. 1979) and Rb-Sr (Papanastassiou et al. 1970) methods range from 3.60 to 3.90 Ga (Beaty and Albee, 1980; Jerde et al., 1994; Snyder et al., 1995; Stöffler et al., 2006). Group A is the youngest basalt group with an age of 3.60 Ga, followed by Group B1 (3.72 Ga), Group B3 (3.76 Ga) and Group B2 (3.90 Ga) (Beaty and Albee, 1980). During later studies, the sample ages were refined as follows: Group A (3.58 G.a), Group B1/B3 (3.70 G.a), Group B2 (3.80 G.a), and Group D (3.85 Ga) (Snyder et al., 1995). Group A is the most abundant rock type in the Apollo 11 sample collection, leading to the interpretation that it represents the surface lithology (Beaty and Albee, 1980). Kuiper et al. (2008) argue that the old Ar-Ar ages need to be updated with the new decay constants to provide corrected ages.

Papanastassiou et al. (1970) proposed that the rock samples collected from the Apollo 11 landing site are from widely different sources and are not genetically related, on the basis of the variations in Rb-Sr isotopes. In particular, they performed Rb-Sr analyses on six crystalline samples from Group A: 10017 (3.59±0.05 Ga), 10057 (3.63±0.002 Ga), 10069 (3.68±0.00 Ga), 10071 (3.68±0.02 Ga) and Group B: 10044 (3.7±0.11 Ga), 10058 (3.63±0.20 Ga) (Papanastassiou et al., 1970). Because this study yielded ages within a narrow range of 3.65±0.06 Ga, they proposed that around this time at least two types of magmatic reservoirs were involved in the formation of lava flows at the Apollo 11 landing site. The flows are described to differ chemically, but not texturally (Papanastassiou et al., 1970). Gaffney et al. (2011) compared ages determined via different isotopic systems (Sm-Nd, Rb-Sr, $^{238}$U -$^{206}$Pb) for sample 10017 in controlled (unshocked and unheated), shocked, and heated states. In the controlled samples, they found ages of 3.633 ± 0.057, 3.678 ± 0.069 and 3.616 ± 0.098 with Sm-Nd, Rb-Sr, and $^{238}$U-$^{206}$Pb respectively (Gaffney et al. 2011). Snape et al., (2016) measured a crystallization age of 3.688±5 Ga for sample 10044 on the basis of Pb-Pb dating. Thus, their age is consistent with the previous work of Papanastassiou and Wasserburg (1971) which produced an Rb-Sr age of 3.71±0.011 and Guggisberg et al., (1979) who determined an Ar-Ar age of 3.71±0.011.

# 7. Calibration of the Lunar Cratering Chronology

The Apollo 11 landing site is one of the important calibration points for the lunar chronology function (Neukum et al., 1983, 2001). We reevaluated the calibration points defined by Apollo 11 samples and the count areas defined by Neukum (1983) with the aim to check and/or improve on the calibration point.

Our study confirms that in light of the radiometric sample ages, spectral information, and measured CSFDs, the Apollo 11 landing site is located in a rather homogeneous mare basalt unit with an age of ~3.60 Ga (e.g., Beaty and Albee, 1980; Stöffler et al., 2006; Snape et al., 2016). After detailed studies of the composition and radiometric ages of the samples. The Group A samples (like sample 10017 Ilmenite basalts) are representative of the mapped mare basalt unit Im3. The measured compositions of $TiO_2$ (~14 wt%) and FeO (~19 wt%) of the samples (Mayer, 2012) coincide with ones measured with Clementine data in unit Im3. Therefore, we plot the average age of these surficial basalts, 3.59±0.05 Ga (e.g., Beaty and Albee, 1980; Snyder et al., 1995, Stöffler et al., 2006) for comparison with our newly measured $N(1)$ values of 6.42x10$^{-3}$±0.54km$^{-2}$ and 6.88x10$^{-3}$±0.611km$^{-2}$ on the Neukum (1983) chronology (Fig. 15). The selected age shows the least error and represents the Group A basalts, rather than individual samples.

We also compared our results with the Neukum et al. (2001) chronology, which shows a difference of about 100 Ma in the absolute model ages (AMAs) (see Appendix A7, A8). Neukum et al. (1983, 1994) argued that the production function did not change between the Nectarian and the Copernican epochs. The production function calculated by Neukum et al. (2001) does not show any significant difference from the production function calculated by Neukum (1983), for the diameter range below 1 km (Stöffler et al. 2006).

CSFD measurements of Neukum (1983) yielded two distinct $N(1)$ values of $9\times10^{-3}$ km$^{-2}$ and $6.4\times10^{-3}$ km$^{-2}$ for the Apollo 11 landing site area, which were compared to the radiometric ages of the samples and interpreted as 'old lava flow' and 'young lava flow' respectively. Consequently, Neukum (1983) introduced two calibration points for the Apollo 11 landing site for the lunar chronology (Fig. 15). However, these two calibration points poorly fit the lunar chronology, which was derived from a least squares fit to all existing data points (Fig. 15). Hiesinger et al. (2000) measured an $N(1)$ value of $7.60\times10^{-3}$ km$^{-2}$ and an AMA of 3.63±0.07 Ga for the mare unit T17 (Fig. 12.b), which contains the Apollo 11 landing site. The AMA determined in this work lies well in the range of the determined radiometric ages showing that the CSFDs and ages for various mare units in the Apollo 11 region are internally consistent. The point derived from Hiesinger et al. (2000) T17 unit's $N(1)$ value also fits relatively better with the Neukum et al (1983) chronology curve (Fig. 15), thus supporting the application of the curve.

Recently, Robbins (2014) determined a single $N(1)$ value of $7.67\times10^{-3}$ km$^{-2}$ with LRO WAC data, which is in the same range as Hiesinger et al. (2000) $N(1)$ values. In Fig. 15, we compared the measured $N(1)$ value with the age of "old" lava flows (after Robbins, 2014) i.e., ~3.72 Ga. However, it is possible that the single value determined by Robbin (2014) also represents the younger lava flow unit. The differences in the $N(1)$ values may also result from the selection of a much larger count area, which can potentially cover many geological units.

Marchi (2009) proposed a model production function and recalculated the $N(1)$ of the Mare Tranquillitatis "old" and "young" units (Fig 15). The "old" unit of Marchi (2009) fits much better to Neukum (1983) lunar chronology than the "young" unit. However, most of the previous studies $N(1)$ were compared with general "old" and "young" age group of the samples, We did not observe different lava flows around the landing site and compared the crater retention values of this unit with representative samples Group A ages of ~3.60 Ga.

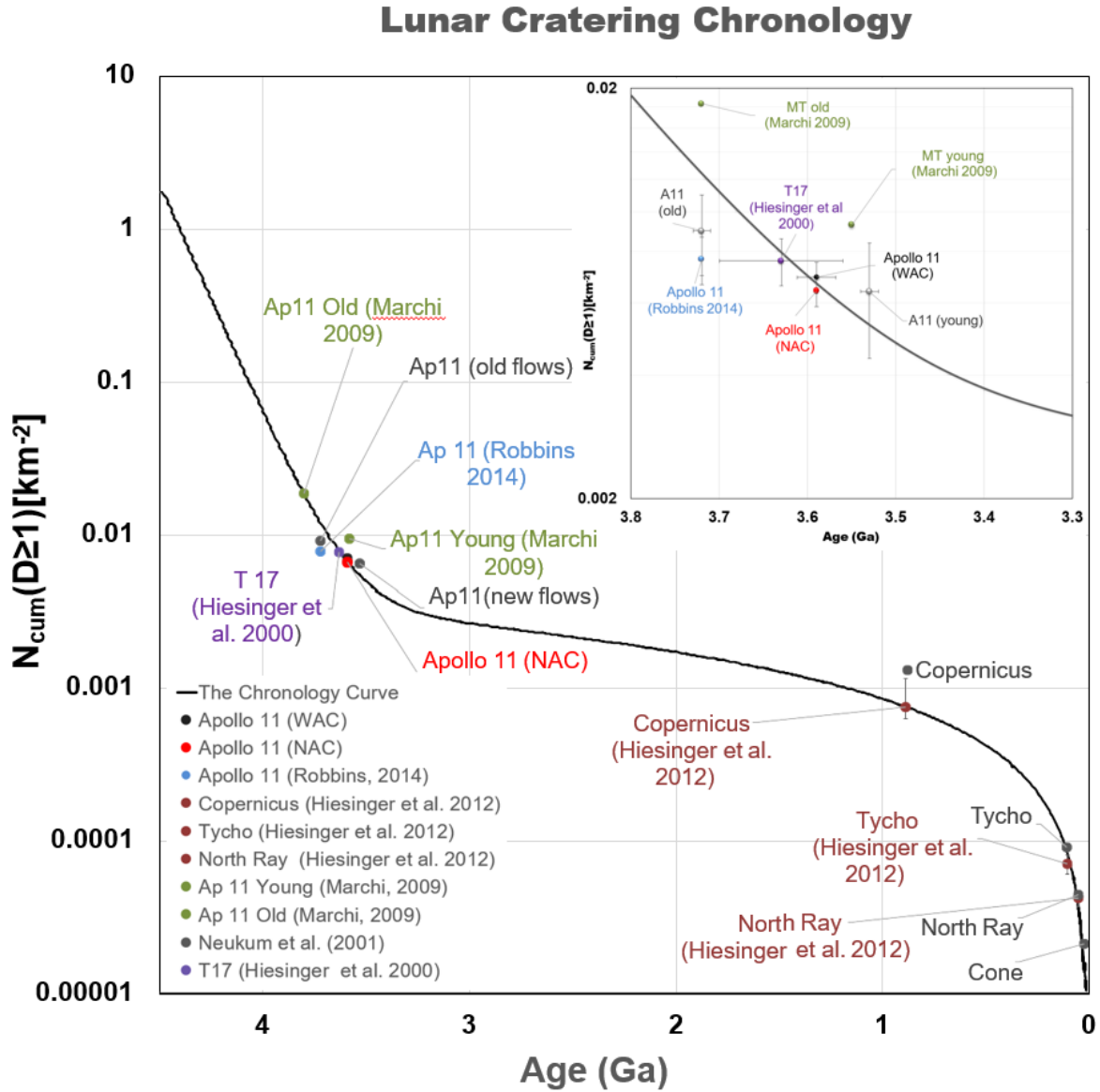

*Figure 15: Lunar cratering chronology curve after Neukum (1983). The gray points represent the values used by Neukum (2001) for the calibration of the curve. The data points in red and black represent our new calculated value of the Apollo 11 landing site, with NAC and WAC data respectively. We also added calibration points from Hiesinger et al. (2000, 2012) in dark red, Marchi (2009) in green, and Robbins (2014) in blue to compare these calibration points on the chronology curve. The inset shows the comparison of our updated values (red and black) to the old values measured by Neukum (2001)(gray), Hiesinger (2000)(purple), Marchi (2009)(green), and Robbins (2014)(blue). We included calculated errors in Table 4.*

# Conclusions

The lunar cratering chronology is a basis for determining ages of unsampled geological units for the terrestrial bodies throughout the Solar System (e.g., Hartmann, 1970; Neukum et al., 1975, 1983, 2001; Hiesinger et al., 2000, 2012; Stöffler et al., 2001, 2006; Marchi, 2009; Robbins, 2014). Thus, it is important to test and improve the calibration points for the chronology function. While there are still uncertainties in determining the exact sources of the samples, detailed geological studies can assist in correlating the samples to likely surface units. In our study, we determined $N$(1) values for geologically homogeneous areas around the Apollo 11 landing site, which were mapped using new high-resolution data. Within our map, we avoided obvious secondary craters in the CSFD measurements. We verified our results by measuring $N$(1) values on differently sized areas and across different diameter ranges on WAC and NAC data. Most basalt samples collected around the landing site represent the surficial high-K basalts, with radiometric age of around 3.60 Ga (e.g., Kramer et al., 1977; Gaffney et al. 2011, Meyer, 2012; Snape et al., 2016; Welsh et al., 2018). However, the radiometric ages still need to be updated with new decay constants (Naumenko-Dèzes et al., 2018) to complete the testing of the Apollo 11 calibration point. Nevertheless, the new CSFDs can be confidently correlated with the surficial high-K group A basalt group for the lunar chronology curve. Neither in the newly mapped area nor on the cumulative crater frequency curve did we observe two mare flows immediately next to the Apollo 11 landing site because we observed a single crater retention age or $N$(1) value for the area around the landing site (Fig. 16). With all the precautions we took to avoid sources of the errors, our new measurements fit well with and support the lunar cratering chronology (Neukum, 1983) for the calibration point of the Apollo 11 landing site, in turn supporting the extension of the chronology to other Solar System bodies.

# Acknowledgments

W. I. and H.H. were supported by the German Research Foundation (Deutsche Forschungsgemeinschaft SFB-TRR170, subproject A2). C. H. vdB. was funded by the German Aerospace Center (Deutsches Zentrum für Luft- und Raumfahrt) project 50OW1504 and as part of a project that has received funding from the European Union's Horizon 2020 research and innovation program under grant agreement Nº776276 (PLANMAP). We thank the reviewers for their helpful and constructive comments and suggestions.

# References

Aldrin, E. E., Jr. N. A. Armstrong and M. Collins (1969), Crew observations: in: Apollo 11 preliminary science report. Houston, Texas, NASA Manned Spacecraft Center. 35-40.

Anderson. D. H., E. E. Anderson, K. Bierman, P. R. Bell, D. D. Bogard, R. Brett, A. L. Burlingame., et al. (1969), Preliminary examination of lunar samples from Apollo 11. Science 165. 3899.

Anderson, J. A., S. C. Sides, D. L. Soltesz, T. L. Sucharski and K. J. Becker (2004), Modernization of the integrated software for imagers and spectrometers. Lunar and Planet. Sci. 35. 2039.

Crater Analysis Techniques Working Group (1979), Standard techniques for presentation and analysis of crater size-frequency data. Icarus 317, 467-474.

Barker, M. K., E. Mazarico, G. A. Neumann, M. T. Zuber, J. Haruyama and D. E. Smith (2016), A new lunar digital model from the Lunar Orbiter Laser Altimeter and SELENE Terrain Camera. Icarus 273, 346-355.

Beaty, D. W., and A. L. Albee (1978), Comparative petrology and possible genetic relations among the Apollo 11 basalts. Proc. Lunar Planet. Sci. Conf. 9: 359-463.

Beaty, D. W., and A. L. Albee (1980), The geology and petrology of the Apollo 11 landing site. Proc. Lunar Planet. Sci. Conf. 11: 23-35.

Blue, J. (1999), Gazetteer of planetary nomenclature. Flagstaff, Ariz, U.S. Geological Survey. http://planetarynames.wr.usgs.gov/.

Compston, W., B. W. Chappell., P. A. Arriens and M. J. Vernon (1970), The chemistry and age of Apollo 11 lunar material. Proc. Apollo 11 Lunar Sci. Conf. 2: 1007-1027.

De Hon, R. A. (1974), Thickness of mare material in the Tranquillitatis and Nectaris basins. Proc. Lunar Planet. Sci. Conf. 1: 53-59.

De Hon, R. A. (2017), A two-basin model for Mare Tranquillitatis. Lunar Planet. Sci. Conf. 48. 2769.

Eugster, O. (1985), The thickness of lava flows in Mare Tranquillitatis and Mare Procellarum. Lunar and Planet. Sci. (16).223-224

Federal Geographic Data Committee (2006), 25-Planetary geology features, FGDC Document number FGDC-STD-013-2016, Appendix A, FGDC Digital Cartographic Standard for Geologic Map Symbolization.

Fortezzo, C. M., and T. M. Hare (2013), Completed digital renovation of the 1:5,000,000 lunar geologic map series. U.S. Geologic Survey, Astrogeology Science Center, Arizona.

Gaffney A. M., L. E. Borg, Y. Asmerom, C. K. Shearer and P. V. Burger (2011), Disturbance of isotope systematics during experimental shock and thermal metamorphism of a lunar basalt with implications for Martian meteorite chronology. Meteor. and Planet. Sci. 46 (1): 35-52

Ganapathy, R., R. R. Keays, J. C. Laul and E. Anders (1970), Trace elements in Apollo 11 lunar rocks: Implications for meteorite influx and origin of moon. Proc. Apollo 11 Lunar Sci. Conf. 2: 1117-1142.

Geiss, J., P. Eberhardt., N. Grögler., S. Guggisberg., P. Maurer and A. Stettler (1977), Absolute time scale of lunar mare formation and filling. Philos. Trans. of the Royal Soci. of London. Series A, Math. and Phys. Sci. 285(1327): 151-158.

Giguere, T. A., G. F. Taylor., B. R. Hawke and P. G. Lucey (2000), The titanium contents of lunar mare basalts. Meteor. and Planet. Sci. 35, 1993-200.

Golombek, M. P., and G. E. McGill (1983), Grabens, basin tectonics, and the maximum total expansion of the moon, J. Geophy. Res., 88 (B4), 3563–3578, doi:10.1029/JB088iB04p03563.

Grolier, M. J. (1970), Geologic map of Apollo site 2 (Apollo 11); Part of Sabine D region, southwestern MareTranquillitatis. USGS Map I-619 [ORB II-6 (25)], scale 125000.

Grove, T. L. and D. W. Beaty (1980), Classification, experimental petrology and possible volcanic histories of the Apollo 11 high-K basalts. Proc. Lunar Planet. Sci. Conf. 11: 149-177.

Guggisberg, S., P. Eberhardt, J. Geiss, N. Grögler and A. Stettler (1979), Classification of the Apollo 11 mare basalts according to $Ar^{39}$-$Ar^{40}$ ages and petrological properties. Proc. Lunar Planet. Sci. Conf. 10: 1-39.

Hartmann, W. K. (1970), Lunar cratering chronology. Icarus 13: 299-301.

Hartmann, W. K. and R. W. Gaskell (1997), Planetary cratering 2: Studies of saturation equilibrium. Meteor. & Planet. Sci. 32(1): 109-121.

Heiken, G., D. Vaniman and B.M. French (1991), Lunar sourcebook: A user's guide to the moon. USA, Cambridge University Press.

Henriksen, M. R., M. R. Manheim, K. N , K.N. Burns, P. Seymour, E. J. Speyerer, A. Derana, A. K. Boyda, E. Howington-Kraus, M. R. Rosiek, B. A. Archinal, and M. S. Robinson (2016), Extracting accurate and precise topography from LROC narrow angle camera stereo observations. Icarus, In Press. doi: 10.1016/j.icarus.2016.05.012

Hess, W. N., and A. J. Calio (1969), Summary of scientific results, in: Apollo 11 preliminary science report. Houston, Texas, NASA Manned Spacecraft Center. 1.

Hiesinger, H., R. Jaumann, G. Neukum and J.W. Head (2000), Ages of mare basalts on the lunar nearside. J. Geophy. Res: Planets 105(E12): 29, 239-229, 275.

Hiesinger, H., J. W. Head, U. Wolf, R. Jaumann and G. Neukum (2002), Lunar mare basalt flow units: Thicknesses determined from crater size-frequency distributions. Geophy. Res. Lett 29(8): 89, 81-89, 84.

Hiesinger, H., J. W. Head, U. Wolf, R. Jaumann and G. Neukum (2003), Ages and stratigraphy of mare basalts in Oceanus Procellarum, Mare Nubium, Mare Cognitum, and Mare Insularum. J. Geophy. Res: Planets 108(E7): 1,1- 1,26.

Hiesinger, H., C. H. van der Bogert, J. H. Pasckert, L. Funcke, L. Giacomini, L.R. Ostrach and M. S. Robinson (2012), How old are young lunar craters? J. Geophy. Res: Planets 117(E12): E00H10.

Hiesinger, H., I. Simon, C. H. van der Bogert, M. S. Robinson and J. B. Plescia (2015), New crater size-frequency distribution measurements for Cone crater at the Apollo 14 landing site. Lunar and Planet. Sci. Conf. 46: 1834.

Ivanov, B. A., G. Neukum, Jr. W. F. Bottke and W. K. Hartmann (2002), The comparison of size-frequency distributions of impact craters and asteroids and the planetary cratering rate. Asteroids III. Tucson, University of Arizona Press: 89-101.

James, O. B. and T. L. Wright (1972), Apollo 11 and 12 mare basalts and gabbros: Classification, compositional variations, and possible petrogenetic relations. Bulletin of the Geol. Soci. of America 83: 2357-2382.

Jerde, E. A., G. A. Snyder, L. A. Taylor, Y. –G. Liu and R. A. Schmitt (1994), The origin and evolution of lunar high-Ti basalts: Periodic melting of a single source at Mare Tranquillitatis. Geochemica et Cosmochimica Acta 58: 515-527.

Kneissl, T., S. van Gasselt and G. Neukum (2011), Map-projection-independent crater size-frequency determination in GIS environments—New software tool for ArcGIS. Planet. and Space Sci. 59(11-12): 1243-1254.

Kramer, F. E., D. B. Twedell and W. J. A. Walton (1977), Apollo 11 lunar sample information catalogue (revised). NASA Lyndon B. Johnson Space Center.

Kring, D. A. (2015), How robotic probes helped (and may again) humans explore the moon. EOS - Earth and Space Sci. News 96.

Kuiper, K. F., A. Deino, F. J. Hilgen., W. Krijgsman., P. R. Renne., and J. R. Wijbrans (2008). Synchronizing Rock Clocks of Earth History. Science, 320(5875), 500–504. doi:10.1126/science.1154339

McEwen, A. S. and E. B. Bierhaus (2006), The importance of secondary cratering to age constraints on planetary surfaces. Annual Review of Earth and Planetary Sciences 34: 535-567.

Meyer, C (2012), Lunar sample compendium. Astromaterials Res. & Explor. Sci. (ARES). NASA.

Meyer, H. M. and A. K. Boyd (2018), Observations from a new global map of Light Plains from the Lunar Reconnaissance Orbiter Camera, 6th European Lunar Symp. (ELS). Toulouse, France

Michael, G. G. and G. Neukum (2010), Planetary surface dating from crater size–frequency distribution measurements: Partial resurfacing events and statistical age uncertainty. Earth and Planet. Sci. Lett. 294(3-4): 223-229.

Michael, G. G., T. Platz, T. Kneissl and N. Schmedemann (2012), Planetary surface dating from crater size–frequency distribution measurements: Spatial randomness and clustering. Icarus 218: 169-177.

Michael, G. G. (2013), Planetary surface dating from crater size–frequency distribution measurements: Multiple resurfacing episodes and differential isochron fitting. Icarus 226(1): 885-890.

Michael, G. G., T. Kneissl and A. Neesemann (2016), Planetary surface dating from crater size-frequency distribution measurements: Poisson timing analysis. Icarus 277: 279-285.

Morris, E. C. and D. E. Wilhelms (1967), Geologic map of the Julius Caesar quadrangle of the moon. USGS Map I-510. (LAC-60).

Neukum, G., B. König and J. Arkani-Hamed (1974), A study of lunar impact crater size-distributions. The Moon, 12: 201-229.

Neukum, G., B. König, H. Fechtig and D. Storzer (1975), Cratering in the earth-moon system - Consequences for age determination by crater counting. In: Lunar Sci. Conf 6, Proceedings 3: 2597-2620.

Neukum, G. (1983), Meteoritenbombardement und datierung planetarer oberflaechen. NASA, Washington, DC, University of Munich: 1-186.

Neukum, G. and B. A. Ivanov (1994), Crater size distributions and impact probabilities on Earth from lunar, terrestrial-planet and asteroid cratering data. in: Hazards due to comets and asteroids 359-416.

Neukum, G., B. A. Ivanov, and W. K. Hartmann (2001), Cratering records in the inner Solar System in relation to the lunar reference system. Space Sci.Reviews 96(1): 55-86.

Naumenko-Dèzes, M. O., T. F. Nägler, K. Mezger, and I. M. Villa (2018), Constraining the 40 K decay constant with 87 Rb- 87 Sr – 40 K- 40 Ca chronometer intercomparison. Geochimica et cosmochimica acta, 220, pp. 235-247. doi: 10.1016/j.gca.2017.09.041

Papanastassiou, D. A., G. J. Wasserburg and D. S. Burnett (1970), Rb-Sr Ages of lunar rocks from the sea of Tranquillitatis. Earth and Planet. Sci. Lett. 8: 1-19.

Papanastassiou, D. A. and G. J. Wasserburg (1971), Lunar chronology and evolution from Rb-Sr studies of Apollo 11 and 12 samples. Earth and Planet. Sci. Lett. 11: 37-62.

Pasckert, J. H., H. Hiesinger, and C. H. van der Bogert (2015), Small-scale lunar farside volcanism. Icarus 257: 336-354.

Pieters, C. M., M. I. Staid, E. M. Fischer, S. Tompkins and G. He (1994), A sharper view of impact craters from Clementine data. Science 266(5192): 1844-1848.

Rajmon, D. and P. Spudis (2004), Distribution and stratigraphy of basaltic units in Maria Tranquillitatis and Fecunditatis: A Clementine perspective. Meteor. & Planet. Sci. 39(10): 1699–1720.

Ringwood, A. E. and E. Essene (1970), Petrogenesis of Apollo 11 basalts, internal constitution and origin of the moon. Proc. Apollo 11 Lunar Sci. Conf. 1: 769-799

Robbins, S. J. (2014), New crater calibrations for the lunar crater-age chronology. Earth and Planet. Sci. Lett. 403: 188-198.

Robinson, M. S., S. M. Brylow, M. Tschimmel, D. Humm, S.J. Lawrence, P.C. Thomas, B.W. Denevi, et al. (2010), Lunar reconnaissance orbiter camera (LROC) instrument overview. Space Sci. Reviews 150(1): 81-124.

Scholten, F., J. Oberst, K.-D. Matz, T. Roatsch, M. Wählisch, E. J. Speyerer and M. S. Robinson (2012), GLD100: The near-global lunar 100 m raster DTM from LROC WAC stereo image data. J. Geophy. Res.: Planets 117(E12): 2156-2202.

Shoemaker, E. M, N. G. Bailey, R. M. Batson, D. H. Dahlem, T. H. Foss, et al. (1969), Geological setting of the lunar samples returned by the Apollo 11 mission, in: Apollo 11 preliminary science report. Houston, Texas, NASA Manned Spacecraft Center. 41-84.

Short, N. M. (1975), Planetary Geology. Englewood Cliffs, New Jersey, Prentice-Hall, Inc.

Smith, J. V, A. T. Anderson, R. C. Newton, E. J. Olsen, P. J. Wyllie, et al. (1970), Petrological history of the moon inferred from petrography, mineralogy, and petrogenesis of Apollo 11 rocks. Proc. Apollo 11 Lunar Sci. Conf. 1: 897-925.

Snape, J. F, A. A. Nemchin, J. J. Bellucci, M. J. Whitehouse, R. Tartèse, et.al. (2016), Lunar basalt chronology, mantle differentiation and implications for determining the age of the moon. Earth and Planet. Sci. Lett. 451: 149-158.

Snyder, G. A, C. M. Hall, L. A. Taylor and A. N. Halliday (1995), 40 Ar/ 39 Ar ages of Apollo 11 group D basalts: Evidence of High-Ti volcanism in the Nectaris Basin and a probable 2.0 Ga age for crater Theophilus? Lunar and Planet. Sci. Conf 26. 1329.

Snyder, G. A, C. M. Hall, D.C. Lee, L. A. Taylor and A. N. Halliday et al. (1996), Earliest high-Ti volcanism on the moon: 40Ar-39Ar, Sm-Nd, and Rb-Sr isotopic studies of Group D basalts from the Apollo 11 landing site. Meteor. & Planet. Sci. 31: 328-334.

Staid, M. I., C.M. Pieters and J. W. Head (1996), Mare Tranquillitatis: Basalt emplacement history and relation to lunar samples. J. Geophy. Res.: Planets 101(E10): 23213-23228.

Stöffler, D., and G. Ryder (2001), Stratigraphy and isotope ages of lunar geologic units: Chronological standard for the inner solar system. Chronology and Evolution of Mars. Netherlands, Springer. 12: 9-54.

Stöffler, D., G. Ryder, B. A. Ivanov, A. Artemieva, M. J. Cintala and R. A. F. Grieve (2006), Cratering history and lunar chronology. Reviews in Mineral. & Geochem. 60: 519-596.

van der Bogert, C. H., G. Michael, T. Kneissl, H. Hiesinger and J. H. Pasckert (2015), Development of Guidelines for Recommended Lunar CSFD Count Areas Sizes via Analysis of Random CSFDs. Workshop on Issues in Crater Studies and the Dating of Planetary Surfaces: 9023Wilhelms, D. E. (1972), Geologic map of the Taruntius quadrangle of the moon. USGS Map I-722. (LAC-61).

Zuber, M. T, D. E. Smith, M. M. Watkins, S. W. Asmar, A. S. Konopliv, et al. (2013), Gravity field of the moon from the gravity recovery and interior laboratory (GRAIL) mission. Science 339 (668): DOI: 10.1126/science.1231507.

# Appendix

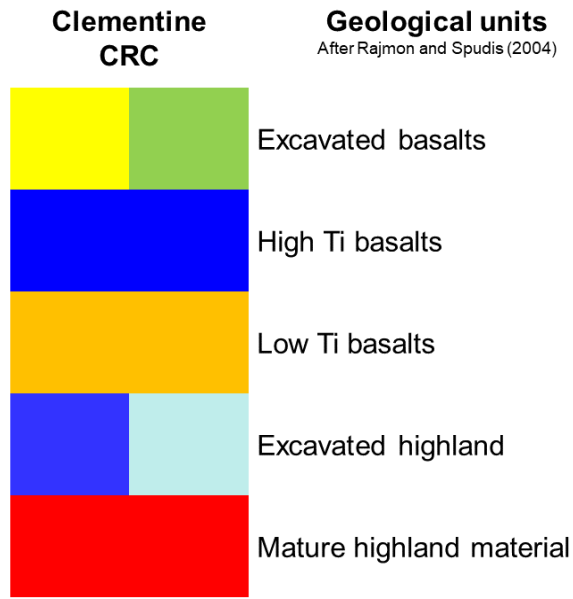

Fig.A1: The color reference chart summarized after Clementine false-color image as proposed by Rajmon and Spudis (2004).

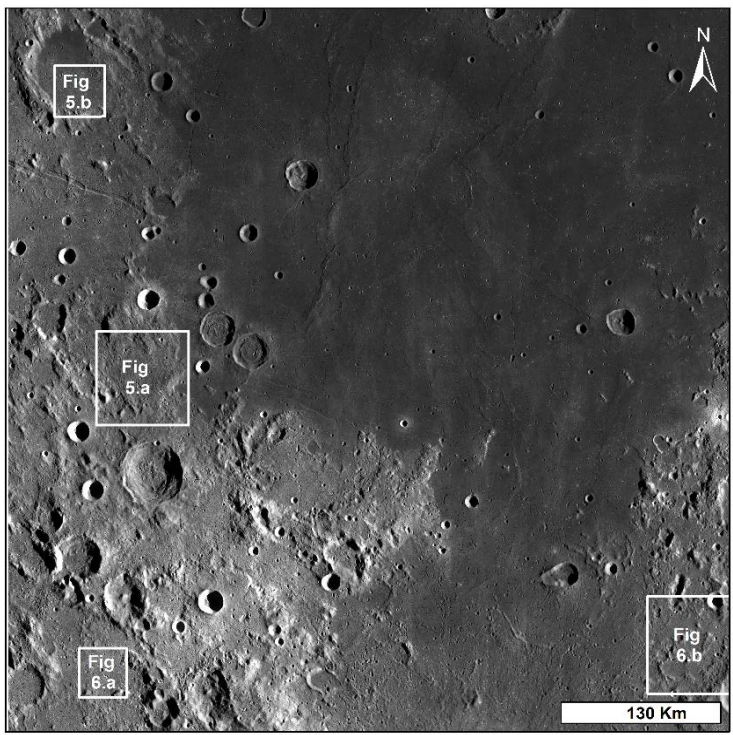

*Fig.A2: Reference map for Figures 5a, 5b, 6a, and 6b, showing examples for different highland units*

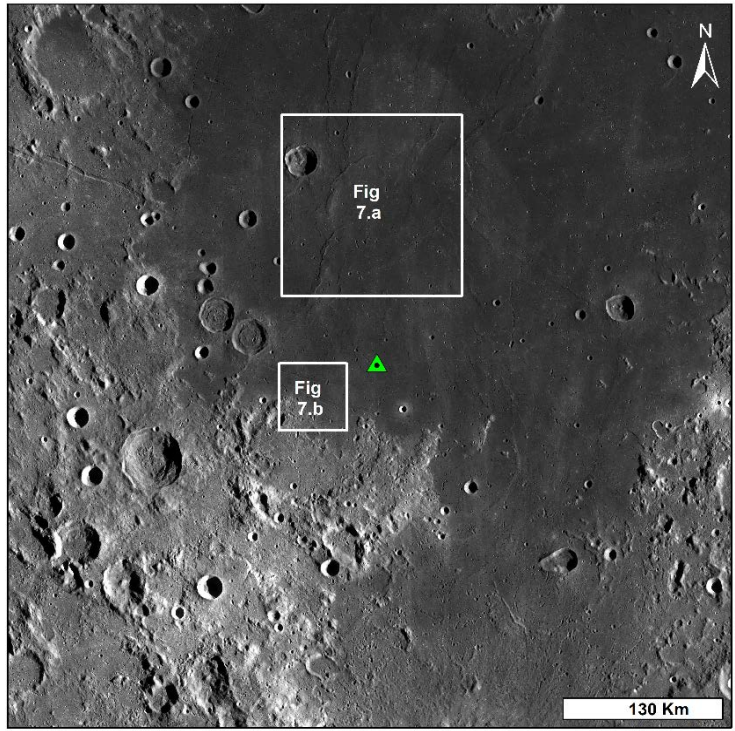

*Fig.A3: Reference map for Figures 7a and 7b, showing examples for the structures present in the mapping area*

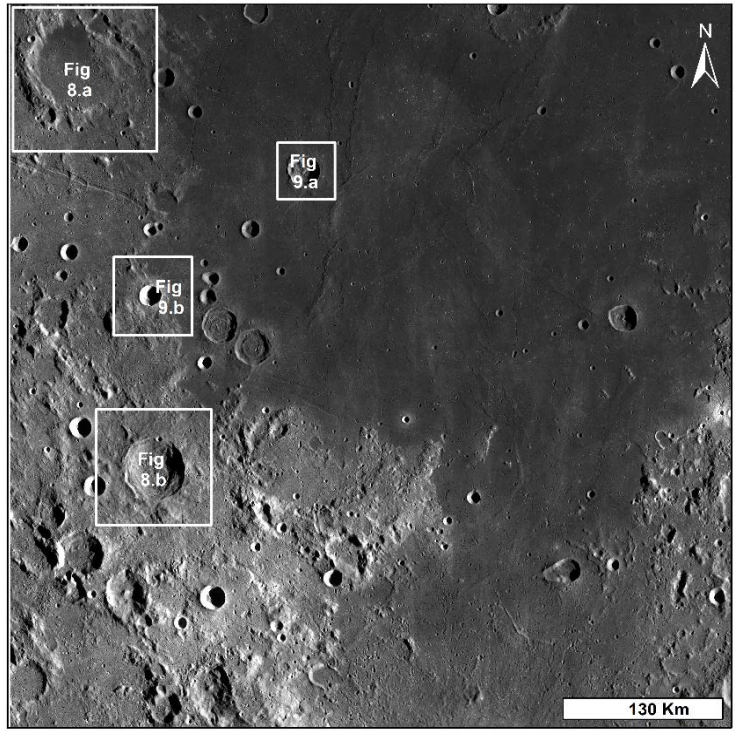

*Fig.A4: Reference map for Figures 8a, 8b, 9a and 9b, showing examples of pre-Imbrian, Imbrian, Erathosthenian and Copernican craters, respectively.*

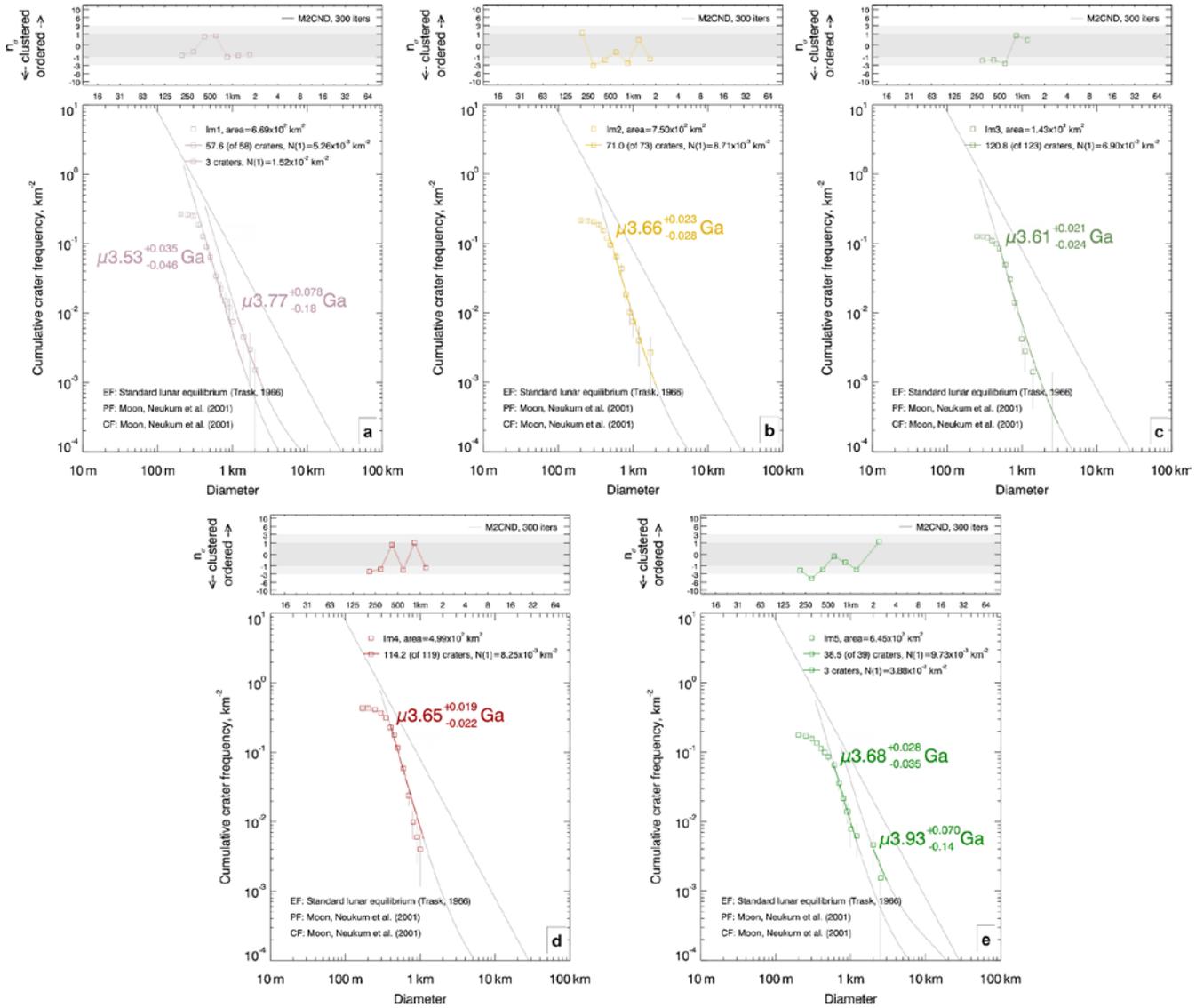

*A5: The CSFD measurements and AMAs of the mare units: (a) Im1 (b) Im2 (c) Im3 (d) Im4 and (e) Im5, shown as cumulative plots with the production function (PF) and chronology function (CF) from Neukum et al. (2001), as well as an equilibrium function (EF) from Trask (1966). These CSFD measurements were made using WAC data. Randomness analysis (top panel) sometimes shows clustering at small diameters, which motivated the exclusion of these small crater diameter ranges from fitting of the CSFDs to determine the N(1) and AMAs values.*

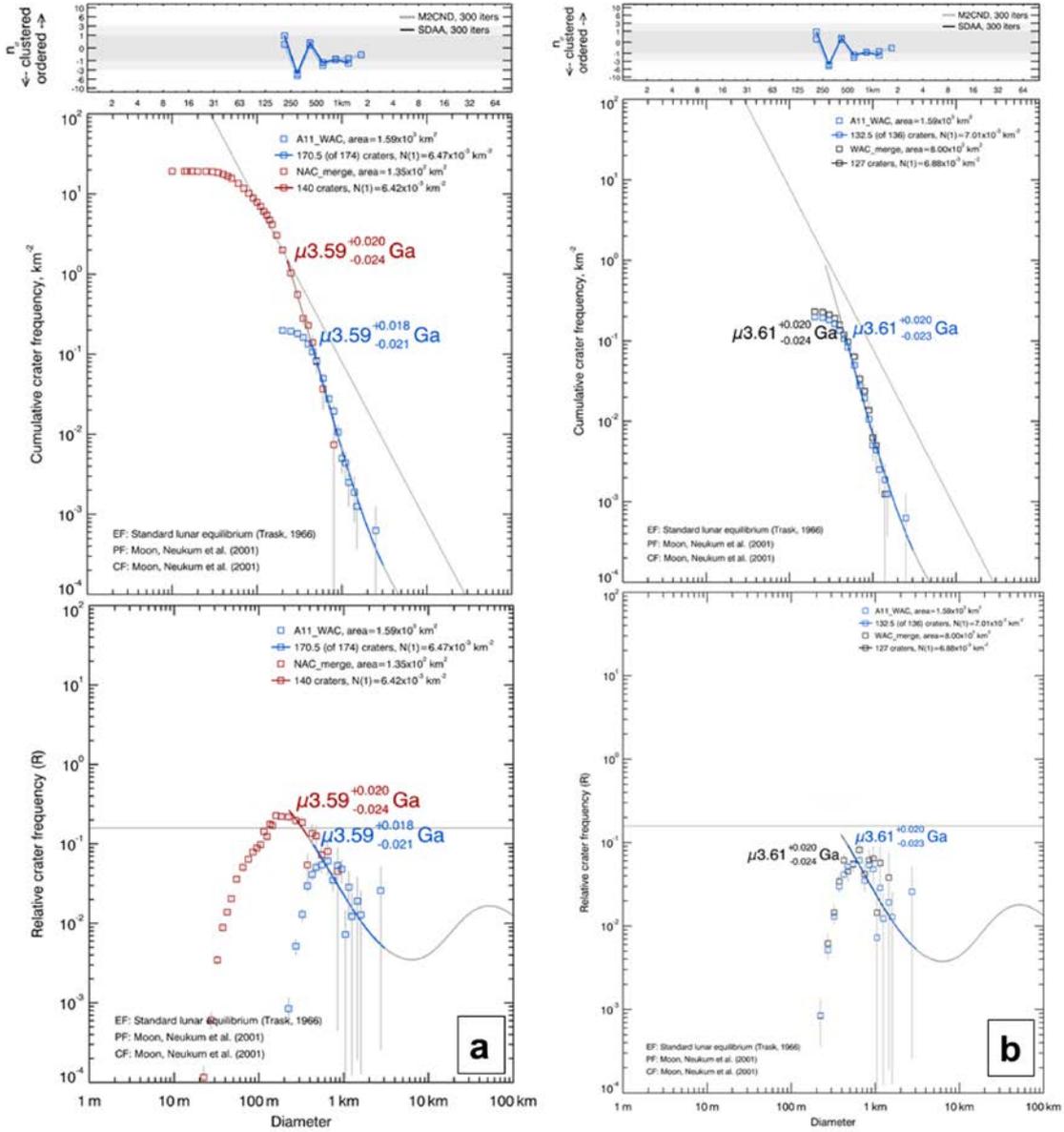

*A6: The CSFD measurements in cumulative and R-Plots: (a) merged N(1) values (red) measured on the NAC data and (b) merged N(1) values (WAC) measured on the WAC data. Both values are compared with the values measured in the area selected by Neukum (1983) (blue).*

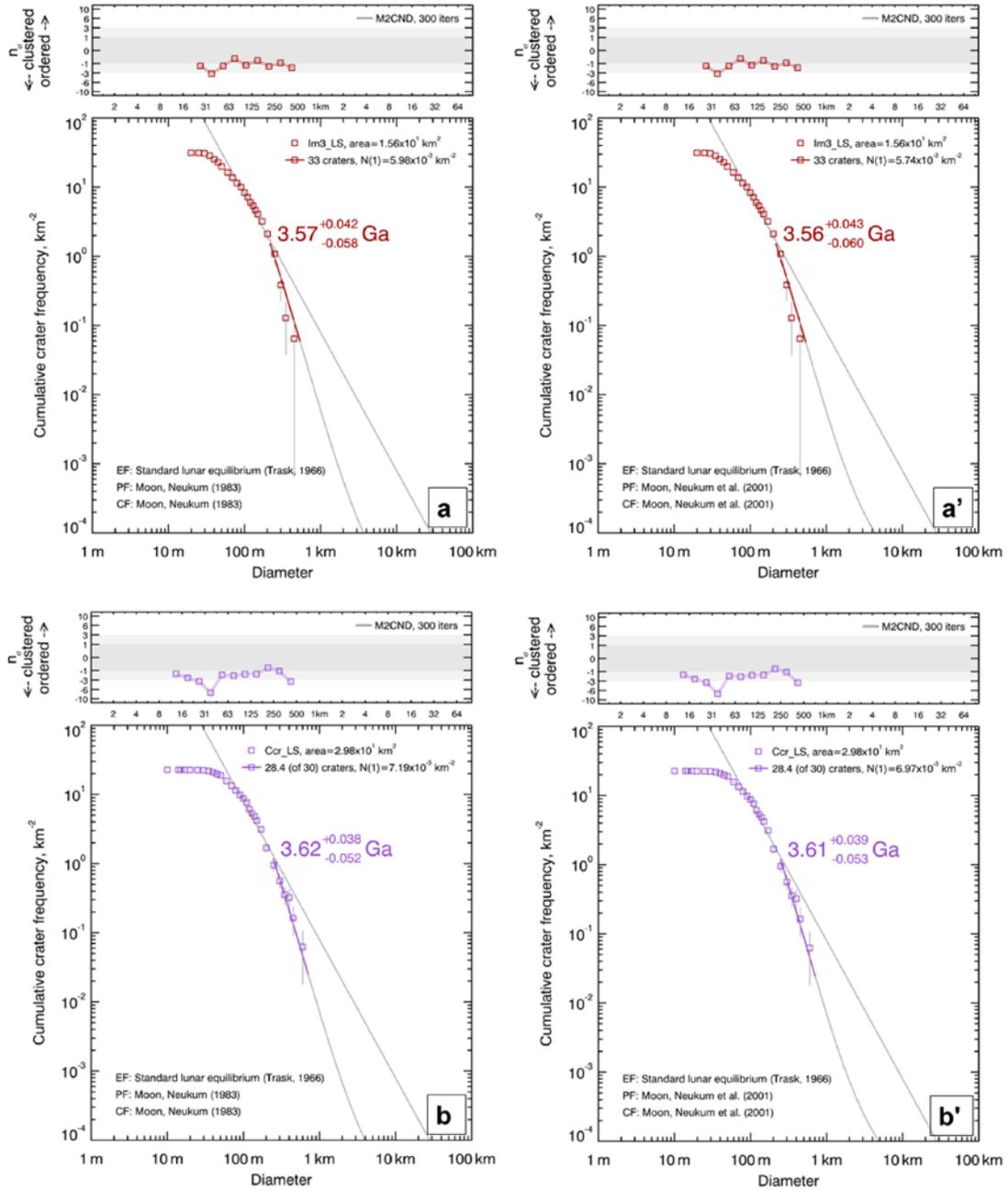

*A7: (a) The CSFD measurements and AMAs of Apollo 11 landing site made on the NAC data: Im3_Ls with PF and CF (1983) and (a') PF and CF (2001). (b)The CSFD measurements and AMAs of the ray material Ccr with PF and CF (1983) and (b') PF and CF (2001). Randomness analyses for both Im3 and Ccr shows the data might have some degree of clustering at small crater diameters. While in Ccr (b,b')bins around 40m shows more clustering.*

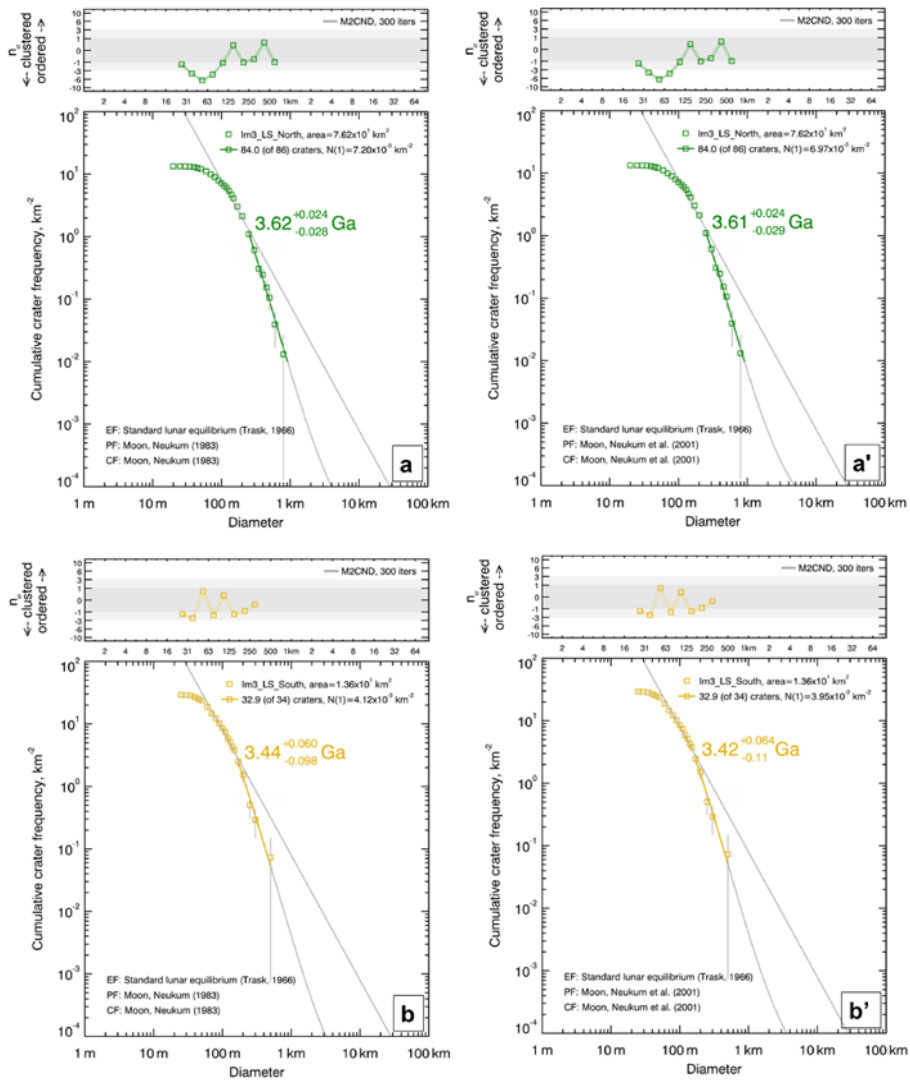

*A8: (a) The NAC CSFD measurements and AMAs of Im3_LS_North with PF and CF (1983) and (a') PF and CF (2001). (b)The CSFD measurements and AMAs of the Im3_LS_South with PF and CF (1983) and PF and CF (2001). Randomness analyses for Im3_LS_North (a,a') shows contamination by secondaries at around 125m, while for Im3_LS_South (b,b') show slight degree of clustering for bin sizes around 40m, 70m and 125m.*